\documentclass[useAMS,usenatbib]{mn2e}
\usepackage{mn2e-breakabs}
\usepackage{graphicx}
\usepackage{subfigure}
\usepackage{times}
\newcommand{\Hunit}{\,{\rm km}\,{\rm s}^{-1}\,{\rm Mpc}^{-1}}

\def\la{\mathrel{\mathpalette\fun <}}
\def\ga{\mathrel{\mathpalette\fun >}}
\def\fun#1#2{\lower3.6pt\vbox{\baselineskip0pt\lineskip.9pt
        \ialign{$\mathsurround=0pt#1\hfill##\hfil$\crcr#2\crcr\sim\crcr}}}

\def\bfs{\mbox{\bf s}}

\newcommand{\be}{\begin{equation}}
\newcommand{\ee}{\end{equation}}
\newcommand{\ba}{\begin{eqnarray}}
\newcommand{\ea}{\end{eqnarray}}
\newcommand{\simgt}{\,\hbox{\lower0.6ex\hbox{$\sim$}\llap{\raise0.6ex\hbox{$>$}}}\,}
\newcommand{\simlt}{\,\hbox{\lower0.6ex\hbox{$\sim$}\llap{\raise0.6ex\hbox{$<$}}}\,}

\begin{document}

\title[SDSS DR7 LRG Multipole of Correlation Function]
{Using Multipoles of the Correlation Function to Measure $H(z)$, $D_A(z)$, and $\beta(z)$ from
Sloan Digital Sky Survey Luminous Red Galaxies}

\author[Chuang \& Wang]{
  \parbox{\textwidth}{
 Chia-Hsun Chuang,\thanks{E-mail: chia-hsun.chuang@uam.es} \thanks{MultiDark Fellow}
 and Yun Wang
  }
  \vspace*{4pt} \\
$^1$ Instituto de F\'{\i}sica Te\'orica, (UAM/CSIC), Universidad Aut\'onoma de Madrid,  Cantoblanco, E-28049 Madrid, Spain \\
$^2$ Homer L. Dodge Department of Physics \& Astronomy, Univ. of Oklahoma,
                 440 W Brooks St., Norman, OK 73019, U.S.A.\\
}

\date{\today} 

\maketitle

\begin{abstract}

Galaxy clustering data can be used to measure the cosmic
expansion history $H(z)$, the angular-diameter distance $D_A(z)$, and the linear
redshift-space distortion parameter $\beta(z)$.
Here we present a method for using effective multipoles of the galaxy two-point correlation
function ($\hat{\xi}_0(s)$, $\hat{\xi}_2(s)$, $\hat{\xi}_4(s)$, and $\hat{\xi}_6(s)$, with $s$ denoting
the comoving separation) to measure $H(z)$, $D_A(z)$, and $\beta(z)$, and validate it 
using LasDamas mock galaxy catalogs. 
Our definition of effective multipoles explicitly incorporates the discreteness
of measurements, and treats the measured correlation function and its theoretical
model on the same footing.
We find that for the mock data, $\hat{\xi}_0+\hat{\xi}_2+\hat{\xi}_4$ captures 
nearly all the information, and gives significantly stronger constraints on $H(z)$, $D_A(z)$, 
and $\beta(z)$, compared to using only $\hat{\xi}_0+\hat{\xi}_2$. 

We apply our method to the sample of luminous red galaxies (LRGs) from the Sloan Digital Sky Survey 
(SDSS) Data Release 7 (DR7) without assuming a dark energy model or a flat Universe.
We find that $\hat{\xi}_4(s)$ deviates on scales of $s<60\,$Mpc$/h$ from the measurement from
mock data (in contrast to $\hat{\xi}_0(s)$, $\hat{\xi}_2(s)$, and $\hat{\xi}_6(s)$), 
thus we only use $\hat{\xi}_0+\hat{\xi}_2$ for our fiducial constraints. 
We obtain $\{H(0.35),D_A(0.35),\Omega_mh^2,\beta(z)\}$ = $\{79.6_{-8.7}^{+8.3}\Hunit$,
$1057_{-87}^{+88}$Mpc, $0.103\pm0.015$, $0.44\pm0.15\}$ using $\hat{\xi}_0+\hat{\xi}_2$.
We find that $H(0.35)\,r_s(z_d)/c$ and $D_A(0.35)/r_s(z_d)$ (where $r_s(z_d)$ is the sound horizon 
at the drag epoch) are more tightly constrained:
$\{H(0.35)\,r_s(z_d)/c, D_A(0.35)/r_s(z_d)\}$ = $\{0.0437_{-0.0043}^{+0.0041}$,$6.48_{-0.43}^{+0.44}\}$ 
using $\hat{\xi}_0+\hat{\xi}_2$. 

\end{abstract}


\section{Introduction}
The cosmic large-scale structure from galaxy redshift surveys provides a 
powerful probe of dark energy and the cosmological model
that is highly complementary to the cosmic microwave 
background (CMB) \citep{Bennett03}, supernovae (SNe) 
\citep{Riess:1998cb,Perlmutter:1998np}, and weak lensing
\citep{Wittman00,bacon00,kaiser00,vw00}. The scope of galaxy redshift 
surveys has dramatically increased in the last decade. The PSCz 
surveyed $\sim 15,000$ galaxies using the Infrared Astronomical Satellite (IRAS)  
\citep{Saunders:2000af}, the 2dF Galaxy Redshift Survey (2dFGRS) 
obtained 221,414 galaxy redshifts \citep{Colless:2001gk,Colless:2003wz}, 
and the Sloan Digital Sky Survey (SDSS) has collected 
930,000 galaxy spectra in the Seventh Data Release (DR7) \citep{Abazajian:2008wr}.
WiggleZ has collected 240,000 emission-line galaxies at $0.5<z<1$ over 
1000 square degrees \citep{Blake09, Parkinson:2012vd}, and BOSS is surveying 1.5 million 
luminous red galaxies (LRGs) at $0.1<z<0.7$ over 10,000 square degrees \citep{Eisenstein11}.
The BOSS data set has been made publicly available recently in SDSS data release 9 \citep{Anderson:2012sa,Manera:2012sc,Nuza:2012mw,Reid:2012sw,Samushia:2012iq,Tojeiro:2012rp}.
The planned space mission Euclid\footnote{http://www.euclid-emc.org/}
will survey over 60 million emission-line galaxies at $0.7<z<2$ over 15,000 square degrees 
\citep{Cimatti:2008kc,Wang10,RB}.

Large-scale structure data from galaxy redshift surveys can be analyzed using either 
the power spectrum or the correlation function. Although these two methods are 
Fourier transforms of one another, the analysis processes are quite 
different and the results cannot be converted using Fourier transform 
directly because of the finite size of the survey volume. 
The SDSS data have been analyzed using both the power spectrum 
method (see, e.g., \citealt{Tegmark:2003uf,Hutsi:2005qv,Padmanabhan:2006ku,Blake:2006kv,Percival:2007yw,Percival:2009xn,Reid:2009xm,Montesano:2011bp}), 
and the correlation function method (see, e.g., 
\citealt{Eisenstein:2005su,Okumura:2007br,Cabre:2008sz,Martinez:2008iu,Sanchez:2009jq,Kazin:2009cj,Chuang:2010dv,Samushia:2011cs,Padmanabhan:2012hf}). 

The power of galaxy clustering as a dark energy probe
lies in the fact that the Hubble parameter, $H(z)$, and the
angular diameter distance, $D_A(z)$, can in principle
be extracted simultaneously from data through the measurement
of the baryon acoustic oscillation (BAO) scale 
in the radial and transverse directions \citep{BG03,SE03,Wang06}. 
\cite{Okumura:2007br} concluded that SDSS DR3 LRG data were
not sufficient for measuring $H(z)$ and $D_A(z)$;
they derived constraints on cosmological parameters assuming that
dark energy is a cosmological constant. \cite{Cabre:2008sz} measured 
the linear redshift space distortion parameter $\beta$, galaxy bias, 
and $\sigma_8$ from SDSS DR6 LRGs. \cite{Gaztanaga:2008xz} 
obtained a measurement of $H(z)$ by measuring the peak of 
the correlation function along the line of sight. However, \cite{Kazin:2010nd}
showed that the amplitude of the line-of-sight peak 
is consistent with sample variance. 

In our previous paper \citep{Chuang:2011fy}, we presented a method to 
measure $H(z)$ and $D_A(z)$ from the full 2D correlation function of
a sample of SDSS DR7 LRGs \citep{Eisenstein:2001cq} without assuming a dark energy model or a flat Universe. 
It is also the first application which includes the geometric distortion 
(also known as Alcock-Paczynski test, see \cite{Alcock:1979mp}) 
on the galaxy clustering data at large scales. 
We demonstrated the feasibility of extracting $H(z)$ and $D_A(z)$
by applying our method to individual LasDamas mock catalogs which mimic the 
galaxy sample and survey geometry of the observational data we used. 
In this paper, we extend our method by exploring the use of the multipoles of the correlation function
to measure $H(z)$, $D_A(z)$, and $\beta(z)$.
The obvious advantage of using multipoles of the correlation function instead of
the full 2D correlation function is the reduced number of data points used to obtain 
similar amounts of information.
In Section \ref{sec:data}, we introduce the galaxy sample used 
in our study. In Section \ref{sec:method}, we describe the details of our 
method. In Section \ref{sec:results}, we present our results. In Sectioin \ref{sec:test},
we apply some systematic tests to our measurements.
We summarize and conclude in Sec.~\ref{sec:conclusion}.

\section{Data} \label{sec:data}

The SDSS-I/II has observed one-quarter of the
entire sky and performed a redshift survey of galaxies, quasars and
stars in five passbands $u,g,r,i,$ and $z$ with a 2.5m telescope
\citep{Fukugita:1996qt,Gunn:1998vh,Gunn:2006tw}. 
We use the public catalog, the NYU Value-Added Galaxy Catalog
(VAGC) \citep{Blanton:2004aa}, derived from the
SDSS II final public data release, Data Release 7 (DR7)
\citep{Abazajian:2008wr}.
We select our LRG sample from the NYU VAGC with 
the flag $primTarget$ bit mask set to $32$. K-corrections
have been applied to the galaxies with a
fiducial model ($\Lambda$CDM with $\Omega_m=0.3$ and $h=1$), and
the selected galaxies are required to have rest-frame $g$-band absolute
magnitudes $-23.2<M_g<-21.2$ \citep{Blanton:2006kt}. The same 
selection criteria were used in previous papers
\citep{Zehavi:2004zn,Eisenstein:2005su,Okumura:2007br,Kazin:2009cj}. 
The sample we use is referred to as ``DR7full'' in \cite{Kazin:2009cj}.
Our sample includes 87000 LRGs in the redshift range 0.16-0.44.

Spectra cannot be obtained for objects closer than 55 arcsec
within a single spectroscopic tile due to the finite size of the
fibers. To correct for these ``collisions'', the redshift of an object
that failed to be measured would be assigned to be the same as the
nearest successfully observed one. Both fiber 
collision corrections and K-corrections have been made in NYU-VAGC 
\citep{Blanton:2004aa}. The collision corrections applied here are 
different from what has been suggested in \cite{Zehavi:2004zn}. 
However, the effect should be small since we are using relatively large 
scale which are less affected by the collision corrections.

We construct the radial selection function as a cubic spline fit 
to the observed number density histogram with the width $\Delta
z=0.01$. The NYU-VAGC provides the description of the
geometry and completeness of the survey in terms of spherical
polygons. We adopt it as the angular selection function of our
sample. We drop the regions with completeness
below $60\%$ to avoid unobserved plates \citep{Zehavi:2004zn}. The 
Southern Galactic Cap (SGC) region is also dropped 
because it consists of three non-contiguous 
stripes in all, and only half as many mocks are available if we
include the SGC in our analysis.

\section{Methodology} 
\label{sec:method}

In this section, we describe the measurement of the multipoles of the correlation function
from the observational data, construction of the theoretical prediction, 
and the likelihood analysis that leads to constraints on dark energy and 
cosmological parameters.

\subsection{Measuring the Two-Dimensional Two-Point Correlation Function}

We convert the measured redshifts of galaxies to comoving distances 
by assuming a fiducial model, $\Lambda$CDM with $\Omega_m=0.25$. 
We use the two-point correlation function (2PCF) estimator given by 
\cite{Landy:1993yu}:
\begin{equation}
\label{eq:xi_Landy}
\xi(\sigma,\pi) = \frac{DD(\sigma,\pi)-2DR(\sigma,\pi)+RR(\sigma,\pi)}{RR(\sigma,\pi)},
\end{equation}
where $\pi$ is the separation along the line of sight (LOS), $\sigma$ 
is the separation in the plane of the sky, DD, DR, and RR represent the normalized 
data-data,
data-random, and random-random pair counts respectively in a
distance range. The LOS is defined as the direction from the observer to the 
center of a pair. The bin size we use here is
$1 \, h^{-1}$Mpc$\times 1 \,h^{-1}$Mpc. 
The Landy and Szalay estimator has minimal variance for a Poisson
process. Random data are generated with the same radial
and angular selection functions as the real data. One can reduce the shot noise due
to random data by increasing the number of random data. The number
of random data we use is 10 times that of the real data. While
calculating the pair counts, we assign to each data point a radial
weight of $1/[1+n(z)\cdot P_w]$, where $n(z)$ is the radial
selection function and $P_w = 4\cdot 10^4$ $h^{-3}$Mpc$^3$ 
\citep{Eisenstein:2005su}.
The weight function is included to minimize the variance of
clustering measurements for an inhomogeneious sample \citep{Feldman:1993ky}.

\subsection{Theoretical Two-Dimensional Two-Point Correlation Function}
We compute the linear power spectra by using CAMB
\citep{Lewis:1999bs}. To include the effect of non-linear structure
formation on the BAOs, we first calculate the dewiggled power spectrum 
\begin{equation} \label{eq:dewiggle}
P_{dw}(k)=P_{lin}(k)\exp\left(-\frac{k^2}{2k_\star^2}\right)
+P_{nw}(k)\left[1-\exp\left(-\frac{k^2}{2k_\star^2}\right)\right],
\end{equation}
where $P_{lin}(k)$ is
the linear matter power spectrum, $P_{nw}(k)$ is the no-wiggle or pure
CDM power spectrum calculated using Eq.(29) from \cite{Eisenstein:1997ik}, 
and $k_{\star}$ is marginalized over with a flat prior over the range of
0.09 to 0.13.

We then use the software package \emph{halofit} \citep{Smith:2002dz} to compute the 
non-linear matter power spectrum:
\begin{eqnarray} \label{eq:halofit}
r_{halofit}(k) &\equiv& \frac{P_{halofit,nw}(k)}{P_{nw}(k)} \\
P_{nl}(k)&=&P_{dw}(k)r_{halofit}(k),
\end{eqnarray}
where $P_{halofit,nw}(k)$ is the power spectrum obtained by applying halofit 
to the no-wiggle power spectrum, and $P_{nl}(k)$ is the non-linear power spectrum. 
We compute the theoretical real space two-point correlation 
function, $\xi(r)$, by Fourier transforming the non-linear power spectrum
$P_{nl}(k)$. 

In the linear regime (i.e., large scales) and adopting the small-angle approximation
(which is valid on scales of interest), the 2D correlation function in the redshift space can 
be written as 
\citep{Kaiser:1987qv,hamilton1992}
\begin{equation}
 \xi^{\star}(\sigma,\pi)=\xi_0(s)P_0(\mu)+\xi_2(s)P_2(\mu)+\xi_4(s)P_4(\mu),
\label{eq:multipole_exp}
\end{equation}
where $s=\sqrt{\sigma^2+\pi^2}$, 
$\mu$ is the cosine of the angle between $\bfs=(\sigma,\pi)$ and the LOS, and 
$P_l$ are Legendre polynomials. The multipoles of $\xi$ 
could be expressed as
\begin{eqnarray}
 \xi_0(r)&=&\left(1+\frac{2\beta}{3}+\frac{\beta^2}{5}\right)\xi(r),\\
 \xi_2(r)&=&\left(\frac{4\beta}{3}+\frac{4\beta^2}{7}\right)[\xi(r)-\bar{\xi}(r)],\\
 \xi_4(r)&=&\frac{8\beta^2}{35}\left[\xi(r)+\frac{5}{2}\bar{\xi}(r)
 -\frac{7}{2}\overline{\overline{\xi}}(r)\right],
\end{eqnarray}
where $\beta$ is the redshift space distortion parameter and
\begin{eqnarray}
 \bar{\xi}(r)&=&\frac{3}{r^3}\int_0^r\xi(r')r'^2dr',\\
 \overline{\overline{\xi}}(r)&=&\frac{5}{r^5}\int_0^r\xi(r')r'^4dr'.
\end{eqnarray}
Next, we convolve the 2D correlation function with the distribution function of 
random pairwise velocities, $f(v)$, to obtain the final model $\xi(\sigma,\pi)$ 
\citep{peebles1980}
\begin{equation} \label{eq:theory}
 \xi(\sigma,\pi)=\int_{-\infty}^\infty \xi^\star\left(\sigma,\pi-\frac{v}{H(z)a(z)}
 \right)\,f(v)dv,
\end{equation}
where the random motions are represented by an exponential form 
\citep{ratcliffe1998,Landy:2002xg}
\begin{equation}
 f(v)=\frac{1}{\sigma_v\sqrt{2}}\exp\left(-\frac{\sqrt{2}|v|}{\sigma_v}\right),
\end{equation}
where $\sigma_v$ is the pairwise peculiar velocity dispersion.

The parameter set we use to compute the theoretical
correlation function is 
$\{H(z), D_A(z), \beta, \Omega_mh^2, \Omega_bh^2, n_s, \sigma_v, k_\star\}$, where
$\Omega_m$ and $\Omega_b$ are the density fractions of matter and
baryons, $n_s$ is the powerlaw index of the primordial matter power spectrum, 
and $h$ is the dimensionless Hubble
constant ($H_0=100h$ km s$^{-1}$Mpc$^{-1}$). 
We set $h=0.7$ while calculating the non-linear power spectra. On the scales 
we use for comparison with data, the theoretical correlation 
function only depends on cosmic curvature and dark energy through
parameters $H(z)$, $D_A(z)$, and $\beta(z)$,             
assuming that dark energy perturbations are unimportant (valid in the simplest dark energy models).
Thus we are able to extract constraints from data that are independent of a dark energy 
model and cosmic curvature.

\subsection{Effective Multipoles of the Correlation Function}  \label{sec:multipoles}

From Eqs.(\ref{eq:multipole_exp}) and (\ref{eq:theory}), we define 
\ba
\label{eq:multipole_1}
\hat{\xi}_l(s) &\equiv & \int_{-\infty}^{\infty}{\rm d} v f(v)\,
\xi_l\left(\sqrt{\sigma^2+\left[\pi-\frac{v}{H(z)a(z)}\right]^2}\right)\nonumber\\
&=& \frac{2l+1}{2}\int_{-1}^{1}{\rm d}\mu\, \xi(\sigma,\pi) P_l(\mu)\nonumber\\
&=& \frac{2l+1}{2}\int_{0}^{\pi}{\rm d}\theta \, \sqrt{1-\mu^2}\, \xi(\sigma,\pi) P_l(\mu),
\ea
where $\mu=\cos\theta$, and $P_l(\mu)$ is the Legendre Polynomial ($l=$0, 2, 4, and 6 here). 
Note that we are integrating over a spherical shell with radius $s$,
while actual measurements of $\xi(\sigma,\pi)$ are done in discrete bins.
To compare the measured $\xi(\sigma,\pi)$ and its theoretical model on the same
footing, we convert the last integral in Eq.(\ref{eq:multipole_1}) into a sum.
This leads to our definition for the effective multipoles of the correlation function:
\begin{equation}\label{eq:multipole}
 \hat{\xi}_l(s) \equiv \frac{\displaystyle\sum_{s-\frac{\Delta s}{2} < \sqrt{\sigma^2+\pi^2} < s+\frac{\Delta s}{2}}(2l+1)\xi(\sigma,\pi)P_l(\mu)\sqrt{1-\mu^2}}{\mbox{Number of bins used in the numerator}},
\end{equation}
where $\Delta s=5$ $h^{-1}$Mpc in this work, and 
\begin{equation}
\sigma=(n+\frac{1}{2})\mbox{$h^{-1}$Mpc}, n=0,1,2,...
\end{equation}
\begin{equation}
\pi=(m+\frac{1}{2})\mbox{$h^{-1}$Mpc}, m=0,1,2,...
\end{equation}
\begin{equation}
\mu\equiv\frac{\pi}{\sqrt{\sigma^2+\pi^2}}.
\end{equation}

Note that both the measurements and the theoretical predictions for the effective multipoles are computed using
Eq.(\ref{eq:multipole}), with $\xi(\sigma,\pi)$ given by the measured correlation function
(see Eq.(\ref{eq:xi_Landy})) for the measured effective multipoles, and 
Eqs.(\ref{eq:multipole_exp})-(\ref{eq:theory}) for their theoretical predictions.
We do not use the conventional definitions of multipoles to extract parameter constraints
as they use continuous integrals. 
Bias could be introduced if the definitions of multipoles are different between measurements from
data and the theoretical model.

\subsection{Covariance Matrix} \label{sec:covar}

We use the 160 mock catalogs from the LasDamas 
simulations\footnote{http://lss.phy.vanderbilt.edu/lasdamas/} 
(McBride et al., in preparation) 
to estimate the covariance matrix of the observed correlation function. 
LasDamas provides mock catalogs matching SDSS main galaxy and LRG samples.
We use the LRG mock catalogs from the LasDamas gamma release with the same cuts as
the SDSS LRG DR7full sample, $-23.2<M_g<-21.2$ and $0.16<z<0.44$.
We have diluted the mock catalogs to match the radial selection function 
of the observational data by randomly selecting the mock galaxies according to the 
number density of the data sample. We calculate the multipoles of the correlation functions 
of the mock catalogs and construct the covariance matrix as
\begin{equation}
 C_{ij}=\frac{1}{N-1}\sum^N_{k=1}(\bar{X}_i-X_i^k)(\bar{X}_j-X_j^k),
\label{eq:covmat}
\end{equation}
where $N$ is the number of the mock catalogs, $\bar{X}_m$ is the
mean of the $m^{th}$ element of the vector from the mock catalog multipoles, and
$X_m^k$ is the value in the $m^{th}$ elements of the vector from the $k^{th}$ mock
catalog multipoles. The data vector ${\bf X}$ is defined by
\be
{\bf X}=\{\hat{\xi}_0^{(1)}, \hat{\xi}_0^{(2)}, ..., \hat{\xi}_0^{(N)}; 
\hat{\xi}_2^{(1)}, \hat{\xi}_2^{(2)}, ..., \hat{\xi}_2^{(N)};...\},
\label{eq:X}
\ee
where $N$ is the number of data points in each measured multipole; $N=16$ in this work.
The length of the data vector ${\bf X}$ depends on how many multipoles are used. 

\subsection{Likelihood}
The likelihood is taken to be proportional to $\exp(-\chi^2/2)$ \citep{press92}, 
with $\chi^2$ given by
\begin{equation} \label{eq:chi2}
 \chi^2\equiv\sum_{i,j=1}^{N_{X}}\left[X_{th,i}-X_{obs,i}\right]
 C_{ij}^{-1}
 \left[X_{th,j}-X_{obs,j}\right]
\end{equation}
where $N_{X}$ is the length of the vectors $X_{th}$ and $X_{obs}$, 
which represent the theoretical model and the observational data respectively.

As explained in \cite{Chuang:2011fy}, instead of recalculating the observed correlation function 
for different theoretical models, we rescale the theoretical correlation function to avoid rendering 
$\chi^2$ values arbitrary.
The rescaled theoretical correlation function is computed by
\begin{equation} \label{eq:inverse_theory_2d}
 T^{-1}(\xi_{th}(\sigma,\pi))=\xi_{th}
 \left(\frac{D_A(z)}{D_A^{fid}(z)}\sigma,
 \frac{H^{fid}(z)}{H(z)}\pi\right),
\end{equation}
where $\xi_{th}$ is given by eq. (\ref{eq:theory}). Hence $\chi^2$ can be rewritten as
\ba 
\label{eq:chi2_2}
\chi^2 &\equiv&\sum_{i,j=1}^{N_{X}}
 \left\{T^{-1}X_{th,i}-X^{fid}_{obs,i}\right\}
 C_{fid,ij}^{-1} \cdot \nonumber\\
 & & \cdot \left\{T^{-1}X_{th,j}-X_{obs,j}^{fid}\right\},
\ea
where $T^{-1}X_{th}$ is a vector given by eq. (\ref{eq:inverse_theory_2d}) with 
$\xi_{th}$ replaced by its effective multipoles (defined by
eq.\ (\ref{eq:multipole})), and $X^{fid}_{obs}$ is the corresponding vector from observational data measured 
assuming the fiducial model in converting redshifts to distances. See \cite{Chuang:2011fy} for a more detailed
description of our rescaling method.

\subsection{Markov Chain Monte Carlo Likelihood Analysis}

We use CosmoMC in a Markov Chain Monte Carlo
likelihood analysis \citep{Lewis:2002ah}. 
The parameter space that we explore spans the parameter set of
$\{H(0.35)$, $D_A(0.35)$, $\Omega_mh^2$, $\beta$, $\Omega_bh^2$, $n_s$, $\sigma_v$, $k_\star\}$. 
Only $\{H(0.35)$, $D_A(0.35)$, $\Omega_mh^2$, $\beta\}$ are well constrained using
SDSS LRGs alone in the scale range of interest. We marginalize over the other parameters, 
$\{\Omega_bh^2$, $n_s$, $\sigma_v$, $k_\star\}$, with the flat priors, 
$\{(0.01859,0.02657)$, $(0.865,1.059)$, $(0,500)$s$^{-1}$km, $(0.09,0.13)h$Mpc$^{-1}\}$, 
where the flat priors of $\Omega_b h^2$ and $n_s$ are centered on 
the measurements from WMAP7 and has width of $\pm7\sigma_{WMAP}$ (with $\sigma_{WMAP}$ from
\cite{Komatsu:2010fb}). These priors
are wide enough to ensure that CMB constraints are not double counted 
when our results are combined with CMB data \citep{Chuang:2010dv}.
We also marginalize over the amplitude of the galaxy correlation function, effectively
marginalizing over a linear galaxy bias.

\section{Results} \label{sec:results}

\subsection{Measurement of multipoles}

\begin{figure}
\centering
\includegraphics[width=0.8 \columnwidth,clip,angle=-90]{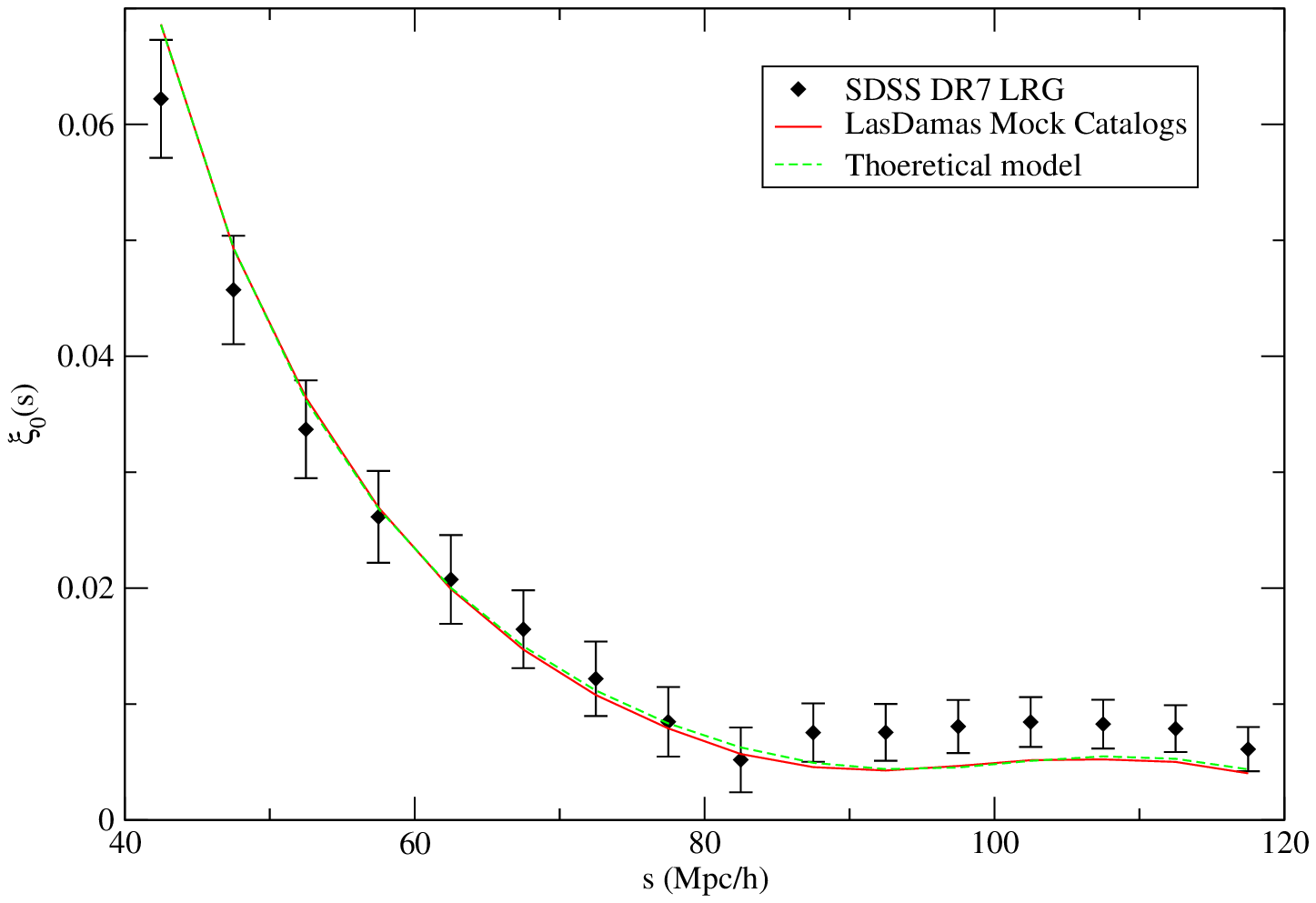}
\caption{Measurement of monopole of the correlation function of the SDSS DR7 LRG (diamond data points), 
compared to the average monopole of the correlation functions of the mock catalogs (solid line) 
and the theoretical model with the input parameters of the simulations (green dashed line). 
The error bars are taken to be the square roots of the diagonal elements of the covariance matrix.}
\label{fig:mono}
\end{figure}

\begin{figure}
\centering
\includegraphics[width=0.8 \columnwidth,clip,angle=-90]{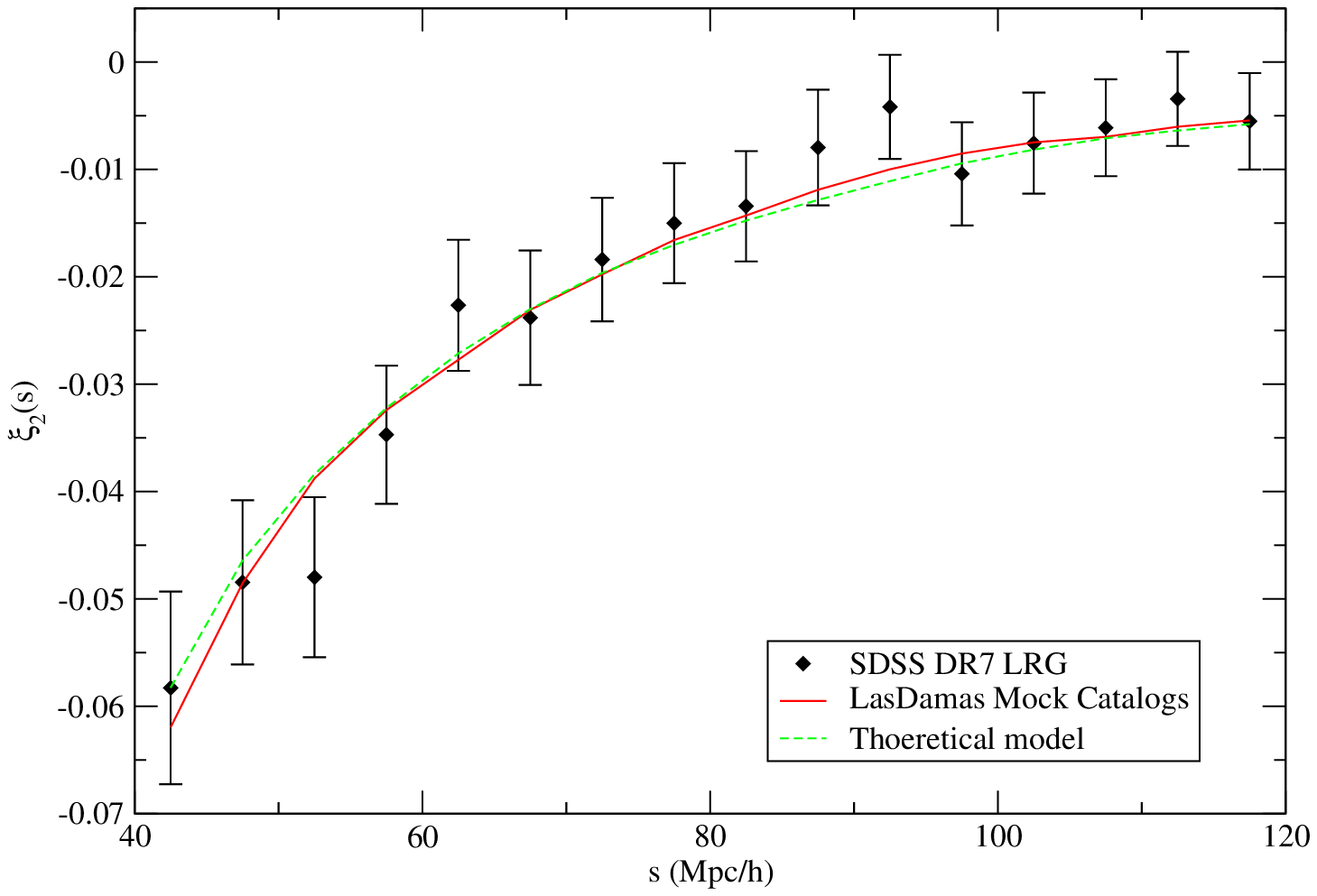}
\caption{Measurement of quadrupole of the correlation function of the SDSS DR7 LRG (diamond data points), 
compared to the average quadrupole of the correlation functions of the mock catalogs (red solid line) 
and the theoretical model with the input parameters of the simulations (green dashed line).  
The error bars are taken to be the square roots of the diagonal elements of the covariance matrix.}
\label{fig:quad}
\end{figure}

\begin{figure}
\centering
\includegraphics[width=0.8 \columnwidth,clip,angle=-90]{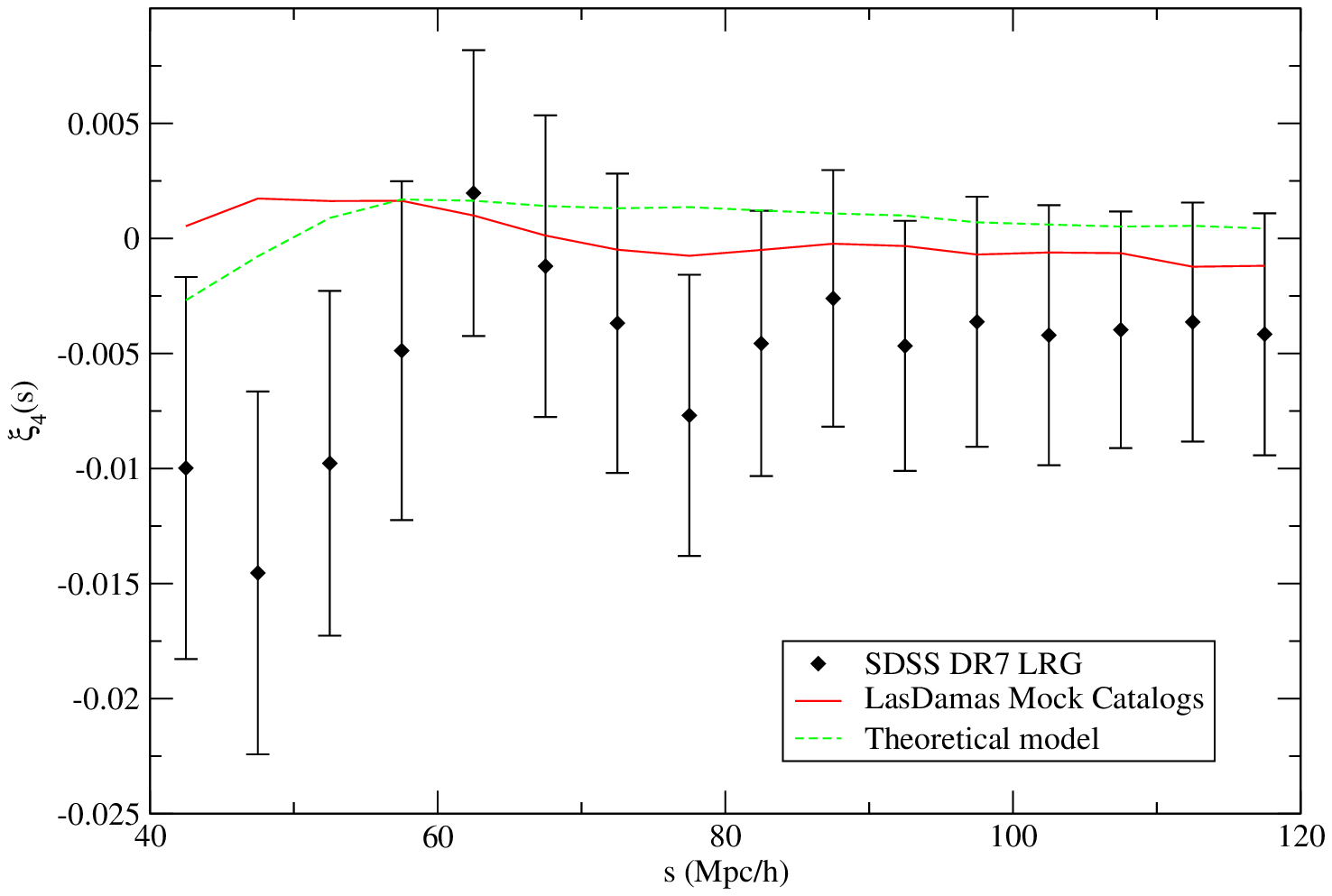}
\caption{Measurement of hexadecapole of the correlation function of the SDSS DR7 LRG (diamond data points), 
compared to the average hexadecapole of the correlation functions of the mock catalogs (red solid line) 
and the theoretical model with the input parameters of the simulations (green dashed line).  
The error bars are taken to be the square roots of the diagonal elements of the covariance matrix.}
\label{fig:four}
\end{figure}

\begin{figure}
\centering
\includegraphics[width=0.8 \columnwidth,clip,angle=-90]{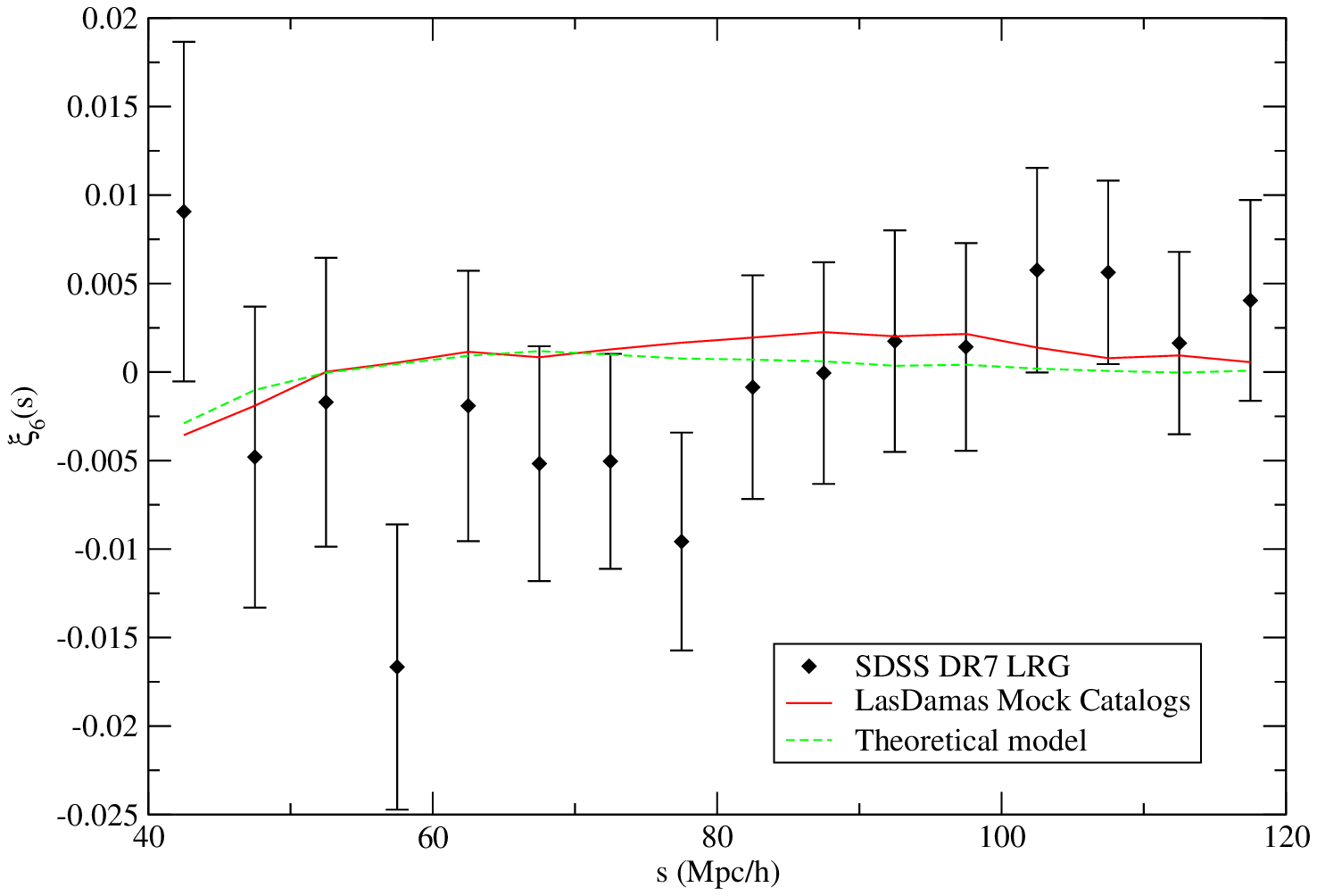}
\caption{Measurement of hexacontatetrapole of the correlation function of the SDSS DR7 LRG (diamond data points), 
compared to the average hexacontatetrapole of the correlation functions of the mock catalogs (red solid line) 
and the theoretical model with the input parameters of the simulations (green dashed line).   
The error bars are taken to be the square roots of the diagonal elements of the covariance matrix.}
\label{fig:six}
\end{figure}

Figs.\ref{fig:mono}, \ref{fig:quad}, \ref{fig:four}, and \ref{fig:six} show the effective 
monopole ($\hat{\xi}_0$), quadrupole ($\hat{\xi}_2$), hexadecapole ($\hat{\xi}_4$), and 
hexacontatetrapole ($\hat{\xi}_6$) measured from SDSS LRGs, 
compared with the average effective multipoles measured from the mock catalogs. 
We use the same scale range as \cite{Chuang:2011fy} ($s=40-120\,h^{-1}$Mpc) for comparison and the bin size 
used is 5 $h^{-1}$Mpc. The data points from the multipoles in the scale range considered are combined to form a 
vector, ${\bf X}$ (see equation(\ref{eq:X})).

\begin{figure}
\centering
\includegraphics[width=0.8 \columnwidth,clip,angle=-90]{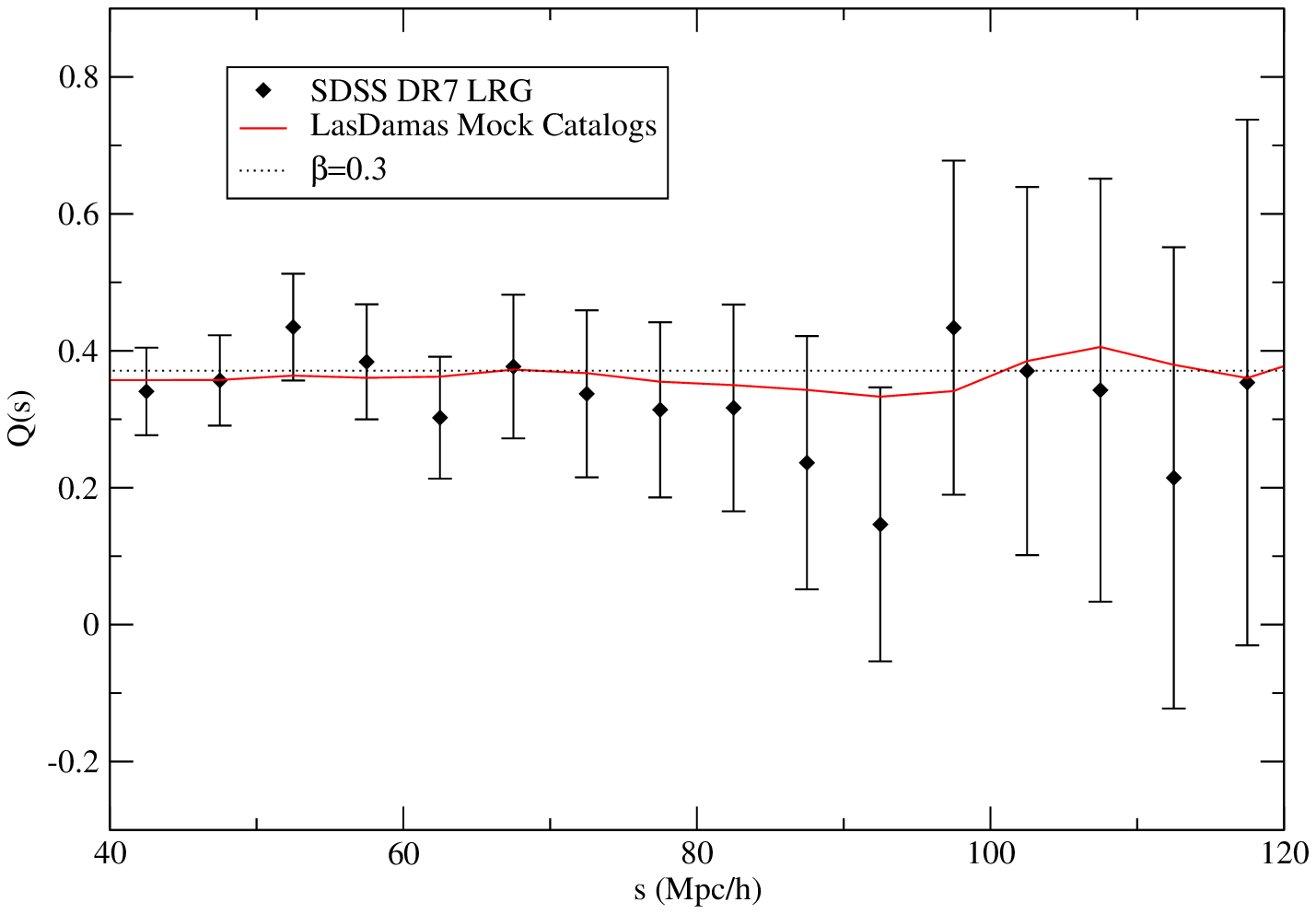}
\caption{Measurement of the normalized quadrupole from the SDSS DR7 LRG (diamond data points), compared to the mean measurement
from the mock catalogs (red solid line). The error bars are taken to be the square roots of the diagonal elements of the covariance matrix.
The green dashed line is the theoretical prediction for $\beta=0.3$ assuming linear power spectrum and small-angle approximation.}
\label{fig:Q_sdss}
\end{figure}

We find that $\xi_4(s)$ deviates on scales of $s<60\,$Mpc$/h$ from the measurement from
mock data (in contrast to $\xi_0(s)$, $\xi_2(s)$, and $\xi_6(s)$).
We note that there are 10 out of 160 mocks which have at least one bin between $40 < s < 55h^{-1}$Mpc for 
which the amplitude of $\xi_4(s)$ is smaller than $-$0.01. 
Therefore, this deviation could be due to the statistical variance.

A frequently used combination of the monopole and the quadrupole is
the normalized quadrupole, defined by
\begin{equation}
 Q(s)=\frac{\xi_2(s)}{\xi_0(s)-(3/s^3)\int_0^s\xi_0(s')s'^2ds'}.
\label{eq:Q}
\end{equation}

For comparison with previous work, we measure the effective normalized quadrupole defined by
\begin{equation}
 \hat{Q}(s) \equiv \frac{\hat{\xi}_2(s)}{\hat{\xi}_0(s)-(3/s^3)\displaystyle\sum_{0<s'\leq s}\hat{\xi}_0(s')s'^2 \Delta s},
\label{eq:Q_eff}
\end{equation}
from SDSS LRGs (see Fig.\ref{fig:Q_sdss}). It is in good agreement with the
expectation from the LasDamas mocks, as well as with previous work by
\cite{Samushia:2011cs}.

\subsection{Measurement of $\{H(0.35)$, $D_A(0.35)$, $\beta(0.35)\}$}

We now present the model independent measurements of the parameters
$\{H(0.35)$, $D_A(0.35)$, $\Omega_m h^2, \beta\}$, obtained by using the 
method described in previous sections. We also present constraints on the derived 
parameters $H(0.35)\,r_s(z_d)/c$ and $D_A(0.35)/r_s(z_d)$, which are more tightly constrained.

\subsubsection{Validation Using Mock Catalogs} 
\label{sec:lasdamas_result}

We first validate our method using mock catalogs.
We have applied it to the first 40 LasDamas mock catalogs 
(which are indexed with 01a-40a)\footnote{We only use 40 instead of 160 mock catalogs because the MCMC is 
computationally expensive. However, the covariance matrix is constructed with 160 mock catalogs.}. Again, we apply the flat and wide priors 
($\pm7\sigma_{WMAP7}$) on $\Omega_b h^2$ and $n_s$, centered on 
the input values of the simulation ($\Omega_b h^2=0.0196$ and $n_s=1$). 

Table \ref{table:mean_lasdamas} shows the means and standard deviations of 
the distributions of our measurements of $\{H(0.35)$, $D_A(0.35)$, $\Omega_m h^2$, $\beta$,
$H(0.35)\,r_s(z_d)/c$, $D_A(0.35)/r_s(z_d)\}$ from 
each monopole + quadrupole ($\hat{\xi}_0+\hat{\xi}_2$) of the LasDamas mock catalogs of the SDSS LRG sample. 
The measurements using
monopole + quadrupole + hexadecapole ($\hat{\xi}_0+\hat{\xi}_2+\hat{\xi}_4$) and 
monopole + quadrupole + hexadecapole + hexacontatetrapole ($\hat{\xi}_0+\hat{\xi}_2+\hat{\xi}_4+\hat{\xi}_6$)
are shown in the same table as well. 
These are consistent with the input parameters, establishing the validity of our method.
In addition, we count the number of the measurements which are outside 1$\sigma$ from the input values of the simulations. 
The measurements include $H(0.35)$, $D_A(0.35)$, and $\Omega_mh^2$ from all three methods, $\hat{\xi}_0+\hat{\xi}_2$, 
$\hat{\xi}_0+\hat{\xi}_2+\hat{\xi}_4$, and  $\hat{\xi}_0+\hat{\xi}_2+\hat{\xi}_4+\hat{\xi}_6$. 
The average percentage is 0.34, close to 0.32, the value we would expect assuming Gaussian distributions.
Note that there is a small difference ($\sim 0.5 \sigma$) between the restored value and 
the input value for $D_A(z)/r_s(z_d)$; it should be possible to remove this by using a
more accurate model for redshift space distortions, e.g., as described in \cite{Reid:2011ar}.
However, applying such models is too computationally expensive in our method.
We will investigate alternative approaches in future work.

While the constraints from using $\hat{\xi}_0+\hat{\xi}_2+\hat{\xi}_4$ are significantly tighter than 
using $\hat{\xi}_0+\hat{\xi}_2$, the constraints from using $\hat{\xi}_0+\hat{\xi}_2+\hat{\xi}_4+\hat{\xi}_6$ 
are nearly the same as that from using $\hat{\xi}_0+\hat{\xi}_2+\hat{\xi}_4$. 
This indicates that $\hat{\xi}_0+\hat{\xi}_2+\hat{\xi}_4$ captures nearly all of the information
that can be extracted from the data given the noise level.
%A flat curve could hardly give constraints. 
Since linear theory predicts that $\hat{\xi}_l=0$ for $l > 4$, it is not surprising
that $\hat{\xi}_0$, $\hat{\xi}_2$, and $\hat{\xi}_4$ capture most of the information 
from the 2D 2PCF.

\begin{figure}
\centering
\includegraphics[width=0.8 \columnwidth,clip,angle=270]{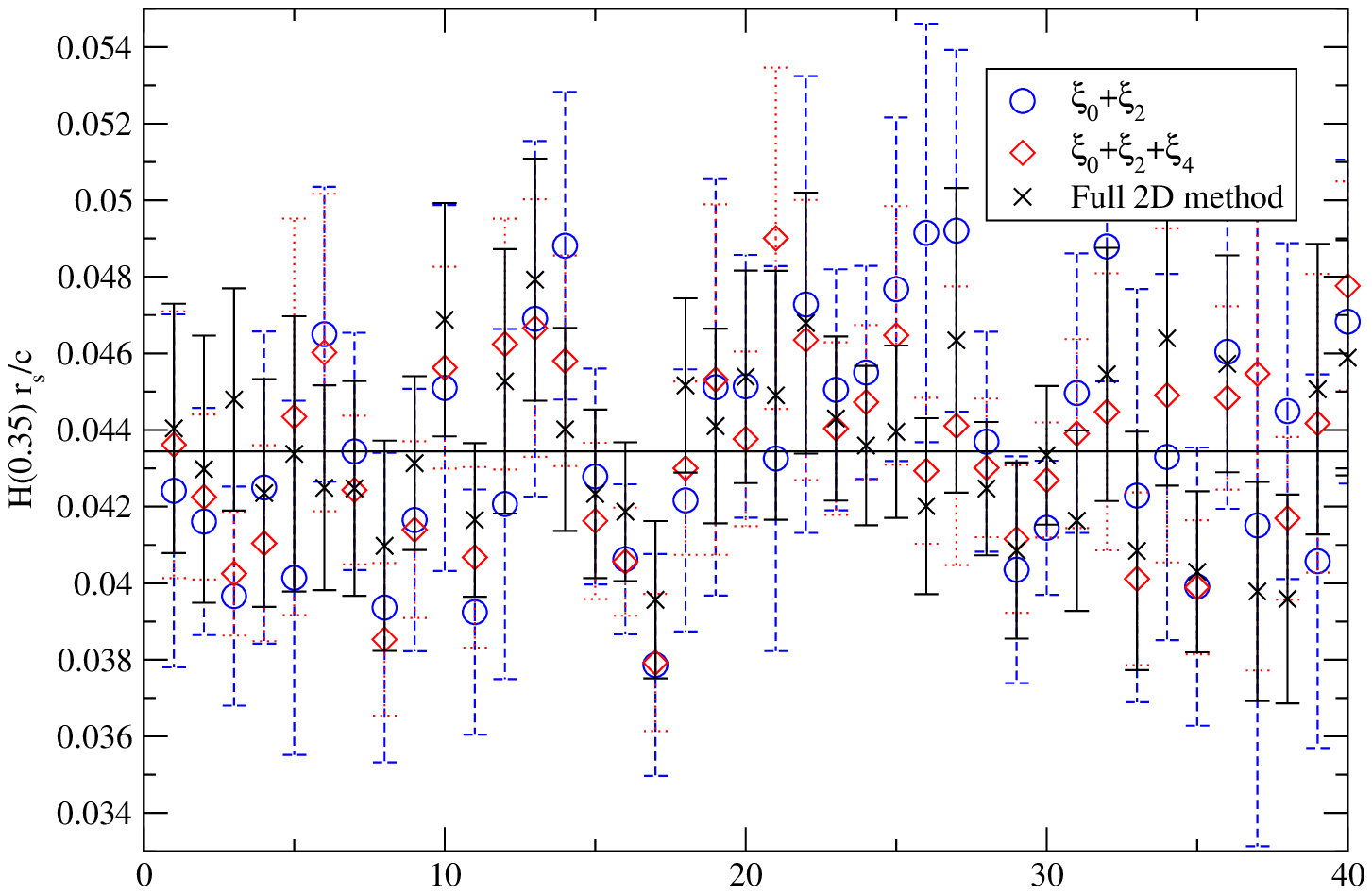}
\caption{Measurements of the means and standard deviation of $H(0.35) \,r_s(z_d)/c$ from 40 individual mock catalogs (indexed as 01a to 40a).
The blue circles show the measurements using $\hat{\xi}_0+\hat{\xi}_2$. The red diamonds show the measurements using $\hat{\xi}_0+\hat{\xi}_2+\hat{\xi}_4$. 
The black crosses show the measurements using full 2D 2PCF method from our previous work. 
The black line shows the theoretical value computed with the input parameters of the simulations.}
\label{fig:Hrs}
\end{figure}

\begin{figure}
\centering
\includegraphics[width=0.8 \columnwidth,clip,angle=270]{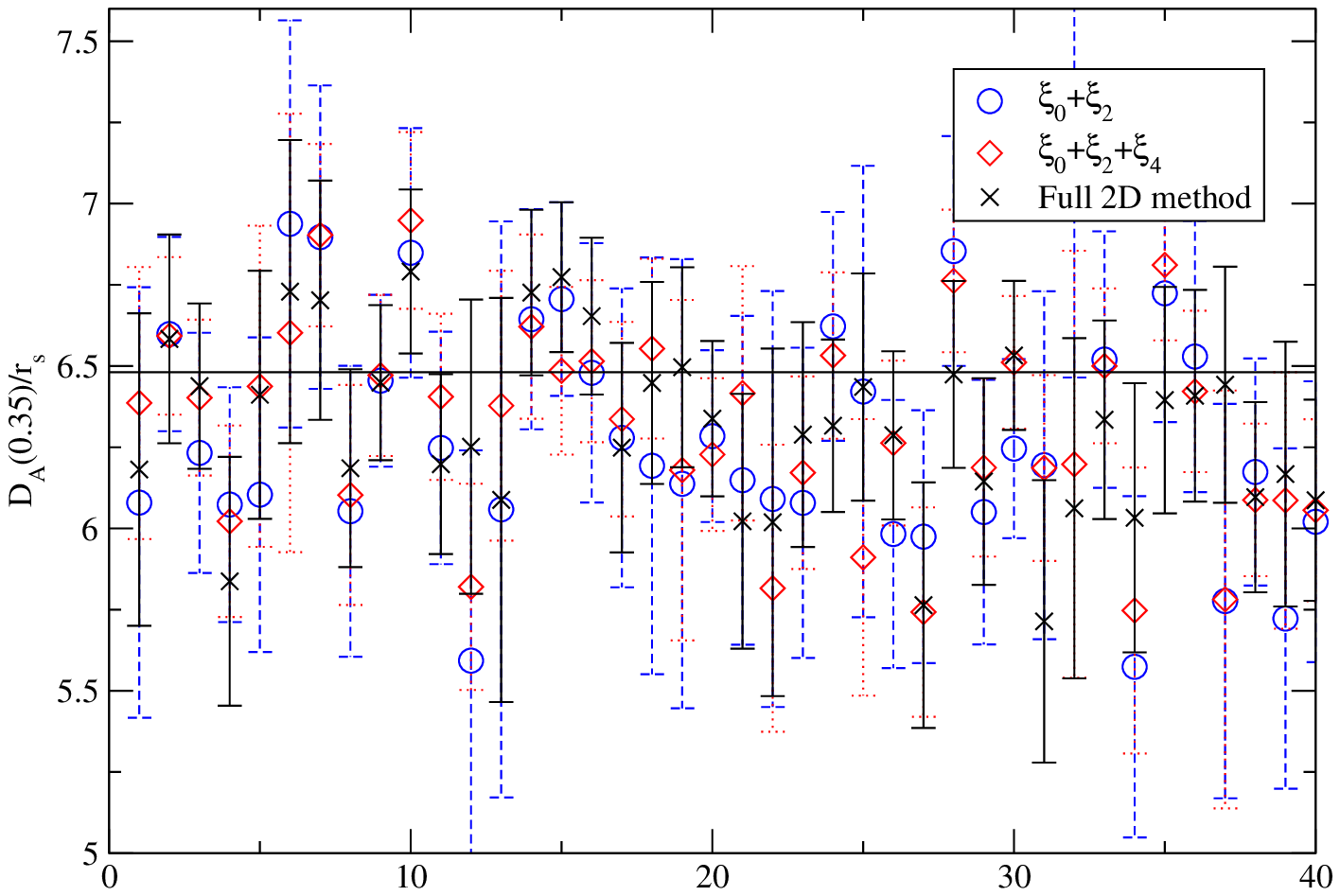}
\caption{Measurements of the means of $D_A(0.35)r_s(z_d)$ from 40 individual mock catalogs (indexed as 01a to 40a).
The blue circles show the measurements using $\hat{\xi}_0+\hat{\xi}_2$. The red diamonds show the measurements using $\hat{\xi}_0+\hat{\xi}_2+\hat{\xi}_4$. 
The black crosses show the measurements using full 2D 2PCF method from our previous work. The black line shows the theoretical value computed with the input parameters of the simulations.
}
\label{fig:rsbyDA}
\end{figure}

\begin{figure}
\centering
\includegraphics[width=0.8 \columnwidth,clip,angle=270]{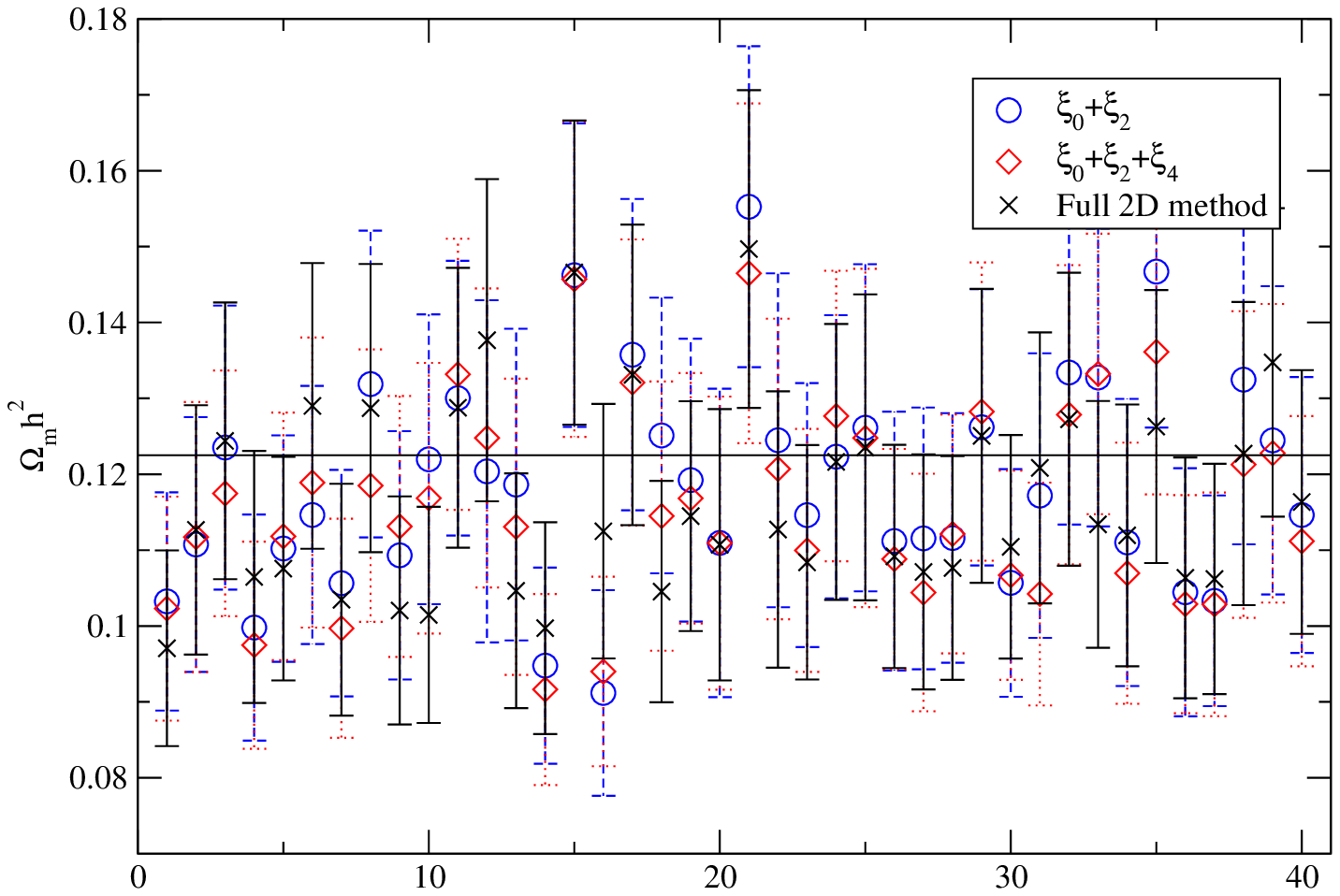}
\caption{Measurements of the means of $\Omega_mh^2$ from 40 individual mock catalogs (indexed as 01a to 40a).
The blue circles show the measurements using $\hat{\xi}_0+\hat{\xi}_2$. The red diamonds show the measurements using $\hat{\xi}_0+\hat{\xi}_2+\hat{\xi}_4$. 
The black crosses show the measurements using full 2D 2PCF method from our previous work. The black line shows the theoretical value computed with the input parameters of the simulations.
}
\label{fig:omh2}
\end{figure}

\begin{figure}
\centering
\includegraphics[width=0.8 \columnwidth,clip,angle=270]{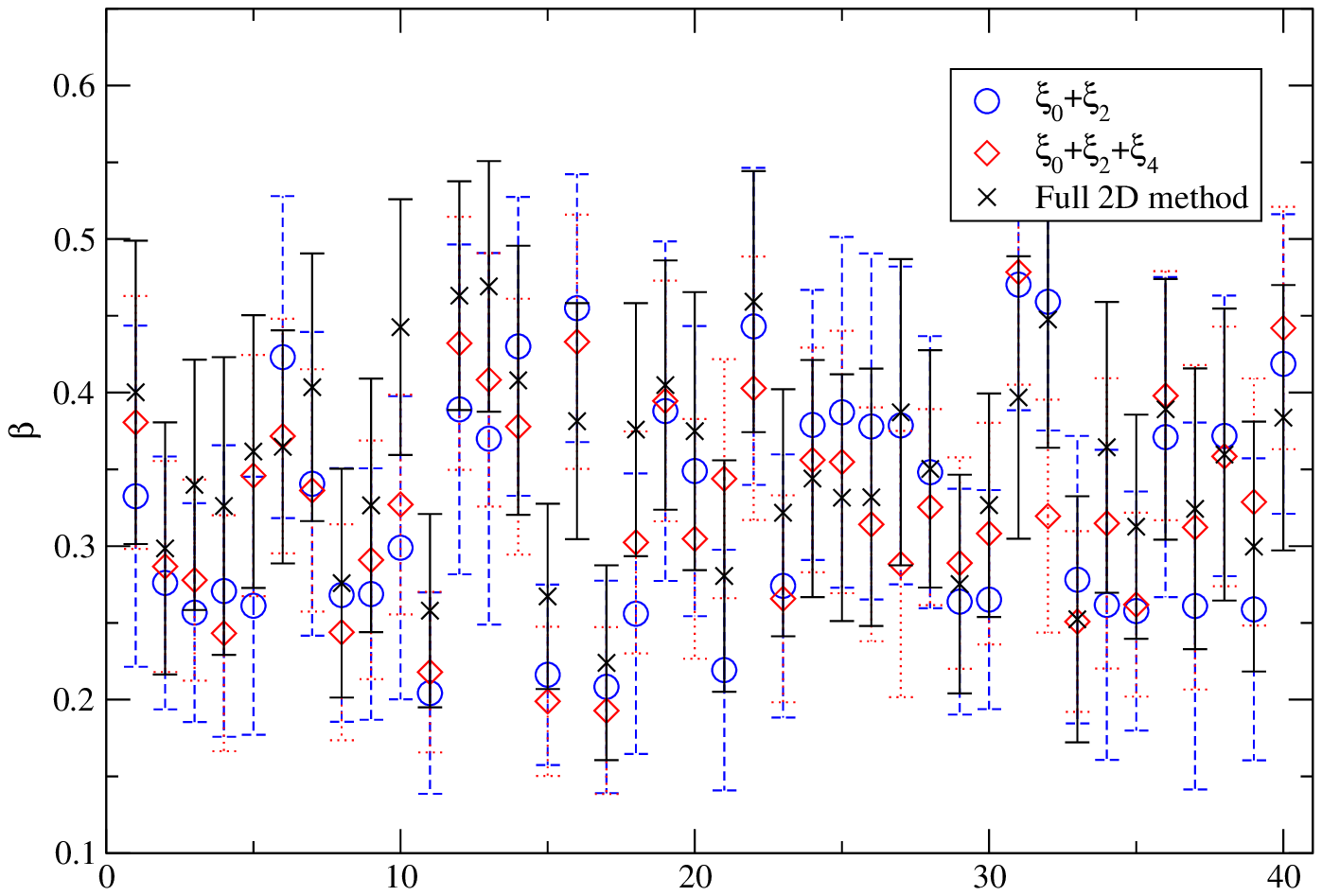}
\caption{Measurements of the means of $\beta$ from 40 individual mock catalogs (indexed as 01a to 40a).
The blue circles show the measurements using $\hat{\xi}_0+\hat{\xi}_2$. The red diamonds show the measurements using $\hat{\xi}_0+\hat{\xi}_2+\hat{\xi}_4$. 
The black crosses show the measurements using full 2D 2PCF method from our previous work.
}
\label{fig:beta}
\end{figure}

In principle, one could obtain better constraints by including more multipoles. However, the tradeoff is 
introducing noise to the covariance matrix which could be a problem, since the number of the mock catalogs 
used to construct the covariance matrix is not big enough.
We also show the measurements of $H(0.35) \,r_s(z_d)/c$, $D_A(0.35)/r_s(z_d)$, $\Omega_mh^2$, and $\beta$ of 
each mock catalog in Fig.\ \ref{fig:Hrs}, Fig.\ \ref{fig:rsbyDA}, Fig.\ \ref{fig:omh2}, and Fig.\ \ref{fig:beta} 
to show the scattering among different mock catalogs and the deviations among different methods. One can see that
the measurements from different methods are consistent for most mock catalogs, but there are still some obvious 
deviations ($>1\sigma$) for a few mock catalogs. 

An important point to note is that since the mock data do not include unknown systematic
effects, the mean values of estimated parameters remain nearly unchanged as more multipoles
measured from data are added to the analysis and the parameter constraints are tightened
with the addition of information.

\begin{table*}
\begin{center}
\begin{tabular}{crrr|r}\hline
&$\hat{\xi}_0+\hat{\xi}_2$  &$\hat{\xi}_0+\hat{\xi}_2+\hat{\xi}_4$ &$\hat{\xi}_0+\hat{\xi}_2+\hat{\xi}_4+\hat{\xi}_6$&input value\\ \hline
$H(0.35)$&\ \ $81.1\pm5.6$ &\ \ $ 80.4\pm $4.9  &\ \ $ 80.3\pm5.1 $ &\ \     81.79 \\
$D_A(0.35)$&\ \  $1017\pm63$ &\ \ $ 1027\pm56 $   &\ \ $ 1021\pm48 $&\ \    1032.8 \\
$\Omega_m h^2$&\ \ $0.119\pm0.014$ &\ \ $ 0.116\pm0.013 $&\ \ $ 0.116\pm0.013 $&\ \  0.1225\\
$\beta$&\ \ $0.325\pm0.076$  &\ \ $ 0.327\pm0.066 $&\ \ $ 0.324\pm0.075 $&\ \  -- \\
\hline
$H(0.35) \,r_s(z_d)/c$&\ \  $0.0436\pm0.0030$   &\ \ $ 0.0435\pm0.0025 $&\ \ $ 0.0434\pm0.0026 $&\ \  0.0434\\
$D_A(0.35)/r_s(z_d)$&\ \ $6.29\pm0.36$ &\ \ $ 6.31\pm0.31 $&\ \ $ 6.28\pm0.26 $&\ \ 6.48\\
\hline
\end{tabular}
\end{center}
\caption{The mean and standard deviation of the distribution of
the measured values of $\{H(0.35)$, $D_A(0.35)$, $\Omega_m h^2$, $\beta$, 
$H(0.35) \,r_s(z_d)/c$, $D_A(0.35)/r_s(z_d)\}$ from 
each $\hat{\xi}_0+\hat{\xi}_2$, $\hat{\xi}_0+\hat{\xi}_2+\hat{\xi}_4$, or $\hat{\xi}_0+\hat{\xi}_2+\hat{\xi}_4+\hat{\xi}_6$ of 40 LasDamas mock catalogs (which are indexed with 01a-40a). 
Our measurements are consistent with the 
input values within 1$\sigma$, where each $\sigma$ is computed from the 40 means measured 
from the 40 mock catalogs. 
The unit of $H$ is $\Hunit$. The unit of $D_A$ and $r_s(z_d)$ is $\rm Mpc$.
} \label{table:mean_lasdamas}
\end{table*}

\begin{figure}
\centering
\includegraphics[width=0.8 \columnwidth,clip,angle=0]{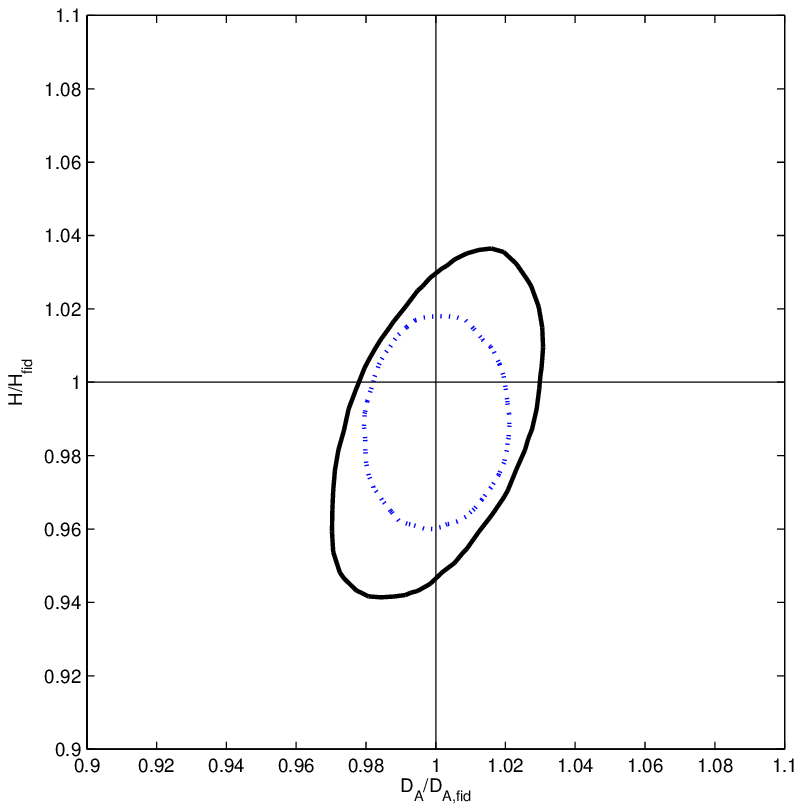}
\caption{2D marginalized contours ($95\%$ C.L.) for $D_A(z)/D_A^{true}$ and $H(z)/H^{true}$ for the comparison with fig. 6 in  
\protect\cite{Kazin:2011xt}. The black solid contour is measured using $\hat{\xi}_0+\hat{\xi}_2$ and the blue dotted contour is measured using $\hat{\xi}_0+\hat{\xi}_2+\hat{\xi}_4$. Our constraints are similar with the results in \protect\cite{Kazin:2011xt}.
}
\label{fig:hda_recov}
\end{figure}

Finally, we compare with the work of 
\cite{Kazin:2011xt}, who measured $H(z)$ and $D_A(z)$ using the average multipoles of the correlation function 
from the LasDamas mock catalogs. They assume a larger survey volume ($\sim12$ times) by dividing the covariance matrix by $\sqrt{160}$. 
They use the average multipoles of the correlation function from the mock catalogs as the theoretical model, which is equivalent to 
fixing $\Omega_mh^2$, $\Omega_bh^2$, and $n_s$ to the input parameters of the simulations. We fix the damping factor, 
$k_\star=0.13$ $h$Mpc$^{-1}$, and the pairwise peculiar velocity dispersion, $\sigma_v=300$ s$^{-1}$km, which give a good 
fit to the average correlation function of the mock catalogs. Corresponding to the bottom panel of fig. 6 in \cite{Kazin:2011xt}, 
we measure the hubble parameter and angular diameter distance by marginalizing over the amplitude of the correlation function 
and the linear redshift space distortion parameter and using the scale range, $s=40-150 h^{-1}$Mpc. Our results are shown in 
Fig. \ref{fig:hda_recov}, and are similar to theirs. The 1-D marginalized uncertainties of $\{H$, $D_A$, $\beta\}$ we 
measure using $\hat{\xi}_0+\hat{\xi}_2+\hat{\xi}_4$ is $\{1.17\%, 0.81\%, 4.45\%\}$ which are similar to their results, 
$\{1.42\%, 0.76\%, 4.95\%\}$ (the numbers are taken from Fig. 7 in \cite{Kazin:2011xt}). They derive the theoretical 
multipoles analytically, instead of using the same definition applied to the observational data. In principle, it could introduce 
biases to the measurements. However, the effect might be minimized since they construct the theoretical model based on 
the measured multipoles from the mock catalogs, which is equivalent to computing the theoretical multipoles with the same 
definition applied to the observational data.

\subsubsection{Measurements from SDSS DR7 LRG} 
\label{sec:sdss_result}

Table \ref{table:mean_sdss_quad} lists the mean, rms variance, and 68\%
confidence level limits of the parameters, $\{H(0.35)$, $D_A(0.35)$, $\Omega_m h^2$, $\beta$, 
$H(0.35) \,r_s(z_d)/c$, $D_A(0.35)/r_s(z_d)\}$,
derived in an MCMC likelihood analysis from the measured $\hat{\xi}_0+\hat{\xi}_2$ of the correlation function
of the SDSS LRG sample.
Table \ref{table:mean_sdss_four} lists the mean, rms variance, and 68\%
confidence level limits of the same parameter set from the measured $\hat{\xi}_0+\hat{\xi}_2+\hat{\xi}_4$ of the correlation function
of the SDSS LRG sample for this parameter set. 
The $\chi^2$ per degree of freedom ($\chi^2/$d.o.f.) is 1.23 for $\hat{\xi}_0+\hat{\xi}_2$ and is 1.06 for $\hat{\xi}_0+\hat{\xi}_2+\hat{\xi}_4$.
These are independent of a dark energy model, and obtained without
assuming a flat Universe.
There are obvious deviations between the cosmlogical constraints obtained from the measured $\hat{\xi}_0+\hat{\xi}_2$ and $\hat{\xi}_0+\hat{\xi}_2+\hat{\xi}_4$
of the correlation function of the SDSS LRG sample, i.e. $\{\Delta\beta=0.10$, $\Delta H(0.35) \,r_s(z_d)/c=0.0037$, $\Delta D_A(0.35)/r_s(z_d)=0.31\}$.
To explore how significant these deviations are, we compute the standard deviations of these differences from Fig. \ref{fig:Hrs}, \ref{fig:rsbyDA}, and \ref{fig:beta} and find
$\{\sigma(\Delta\beta)=0.049$, $\sigma(\Delta H(0.35) \,r_s(z_d)/c)=0.0024$, $\sigma(\Delta D_A(0.35)/r_s(z_d))=0.24\}$. 
One can see that the differences from the measurements are around 1 to 2 $\sigma$.
Thus the deviations between Table \ref{table:mean_sdss_quad} and Table \ref{table:mean_sdss_four} could be due to statistical variance.

Table \ref{table:covar_matrix} gives the normalized covariance matrix
for this parameter set measured using $\hat{\xi}_0+\hat{\xi}_2$.
While the measurement of $\beta$, $0.44\pm0.15$, seems to be higher than what we expect (i.e $\beta=0.325\pm0.076$ from the mock catalogs 
using $\hat{\xi}_0+\hat{\xi}_2$), note that there is a negative correlation between $\beta$ and $\Omega_mh^2$ and 
the correlation coefficient is $-$0.2549. 
Thus the somewhat high $\beta$ value is mildly correlated with the somewhat low $\Omega_m h^2$ value.
In addition, the somewhat high $\beta$ value is actually still statistically consistent with the measurement from the mock catalogs.
The most robust measurements are that of $\{H(0.35) \,r_s(z_d)/c$, $D_A(0.35)/r_s(z_d)\}$,
same as in \cite{Chuang:2011fy}.
These can be used to combine with other data sets and constraining dark energy and cosmological parameters,
see \cite{Wang:2011sb}.

\begin{figure*}
\centering
\includegraphics[width=1 \linewidth,clip]
{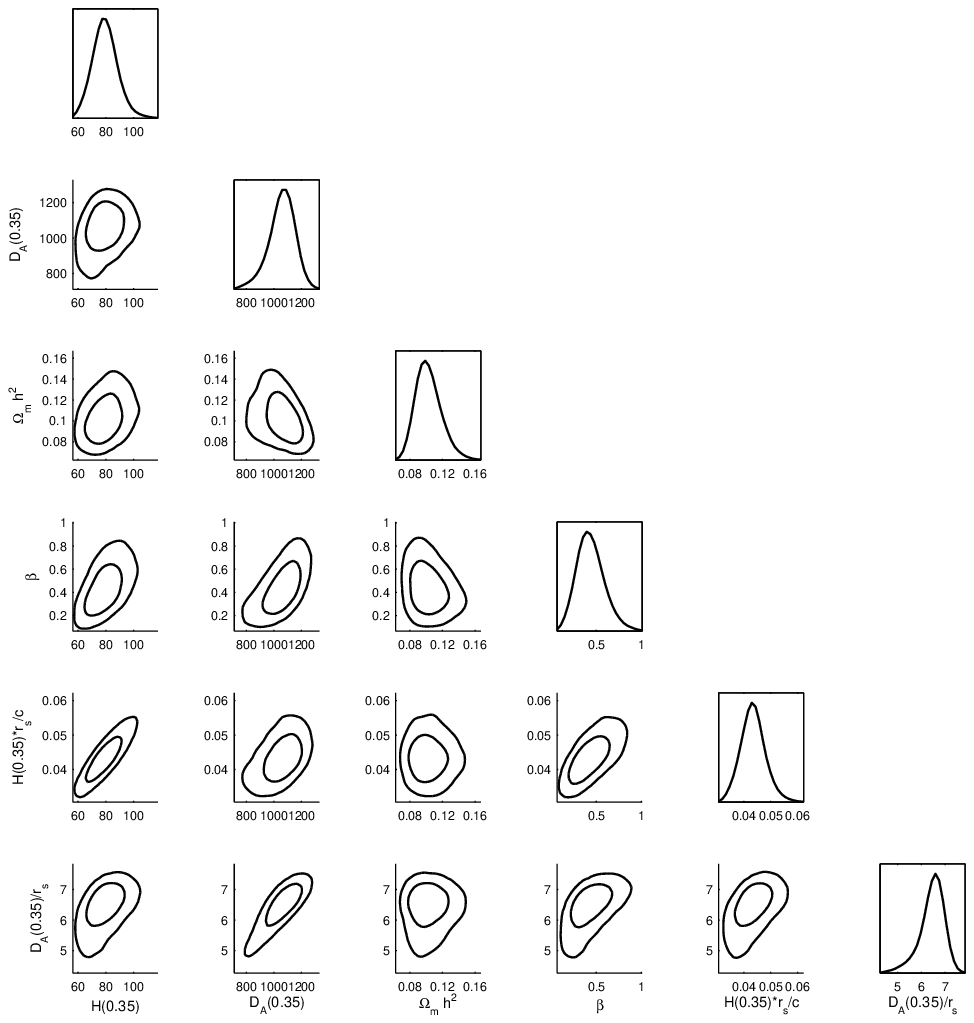}
\caption{2D marginalized
contours ($68\%$ and $95\%$ C.L.) for $\{H(0.35)$, $D_A(0.35)$, $\Omega_m h^2$,  $\beta$,
$H(0.35) \,r_s(z_d)/c$, $D_A(0.35)/r_s(z_d)\}$. The diagonal
  panels represent the marginalized probabilities.
The unit of $H$ is $\Hunit$. The unit of $D_A$ and $r_s(z_d)$ is $\rm Mpc$.}
  \label{fig:sdss_four}
\end{figure*}

Fig.\ \ref{fig:sdss_four} shows one and two-dimensional marginalized contours 
of the parameters, $\{H(0.35)$, $D_A(0.35)$, $\Omega_m h^2$, $\beta$,
$H(0.35) \,r_s(z_d)/c$, $D_A(0.35)/r_s(z_d)\}$,
derived in an MCMC likelihood analysis from the measured $\hat{\xi}_0+\hat{\xi}_2$
of the SDSS LRG sample.

\begin{table}
\begin{center}
\begin{tabular}{crrrr}\hline
&mean &$\sigma$ &lower &upper \\ \hline
$	H(0.35)	$&\ \ 	79.6	&\ \ 	8.8	&\ \ 	70.9	&\ \ 	87.8	\\
$	D_A(0.35)$&\ \ 	1060	&\ \ 	92	&\ \ 	970	&\ \ 	1150	\\
$	\Omega_m h^2	$&\ \ 	0.103	&\ \ 	0.015	&\ \ 	0.088	&\ \ 	0.118	\\
$	\beta	$&\ \ 	0.44	&\ \ 	0.15	&\ \ 	0.29	&\ \ 	0.59	\\
\hline
$	H(0.35) \,r_s(z_d)/c	$&\ \ 	0.0435	&\ \ 	0.0045	&\ \ 	0.0391	&\ \ 	0.0477	\\
$	D_A(0.35)/r_s(z_d)	$&\ \ 	6.44	&\ \ 	0.51	&\ \ 	5.99	&\ \ 	6.90	\\
\hline
\end{tabular}
\end{center}
\caption{
The mean, standard deviation, and the 68\% C.L. bounds of 
$\{H(0.35)$, $D_A(0.35)$, $\Omega_m h^2$, $\beta$,
$H(0.35) \,r_s(z_d)/c$, $D_A(0.35)/r_s(z_d)\}$ from SDSS DR7 LRGs using $\hat{\xi}_0+\hat{\xi}_2$. 
The unit of $H$ is $\Hunit$. The unit of $D_A$ and $r_s(z_d)$ is $\rm Mpc$.
} \label{table:mean_sdss_quad}
\end{table}

\begin{table}
\begin{center}
\begin{tabular}{crrrr}\hline
&mean &$\sigma$ &lower &upper \\ \hline
$	H(0.35)	$&\ \ 	87.3	&\ \ 	6.7	&\ \ 	80.8	&\ \ 	93.7	\\
$	D_A(0.35)$&\ \ 	1095	&\ \ 	59	&\ \ 	1037	&\ \ 	1153	\\
$	\Omega_m h^2	$&\ \ 	0.107	&\ \ 	0.015	&\ \ 	0.093	&\ \ 	0.122	\\
$	\beta	$&\ \ 	0.54	&\ \ 	0.11	&\ \ 	0.44	&\ \ 	0.65	\\
\hline
$	H(0.35) \,r_s(z_d)/c	$&\ \ 	0.0472	&\ \ 	0.0033	&\ \ 	0.0441	&\ \ 	0.0503	\\
$	D_A(0.35)/r_s(z_d)	$&\ \ 	6.75	&\ \ 	0.25	&\ \ 	6.52	&\ \ 	6.98	\\
\hline
\end{tabular}
\end{center}
\caption{
The mean, standard deviation, and the 68\% C.L. bounds of 
$\{H(0.35)$, $D_A(0.35)$, $\Omega_m h^2$, $\beta$,
$H(0.35) \,r_s(z_d)/c$, $D_A(0.35)/r_s(z_d)\}$ from SDSS DR7 LRGs using $\hat{\xi}_0+\hat{\xi}_2+\hat{\xi}_4$. 
The unit of $H$ is $\Hunit$. The unit of $D_A$ and $r_s(z_d)$ is $\rm Mpc$.
} \label{table:mean_sdss_four}
\end{table}

\begin{table*}
\begin{center} 
\begin{tabular}{crrrrrrrr}\hline
       &$H(0.35)$ &$D_A(0.35)$   &$\Omega_mh^2$ &$\beta$ &$H(0.35) \,r_s(z_d)/c$ &$D_A(0.35)/r_s(z_d)$ \\ \hline
$H(0.35)$&\ \ 	1	&\ \ 	0.2669	&\ \ 	0.3529	&\ \ 	0.5802	&\ \ 	0.9259	&\ \ 	0.4832	\\
$D_A(0.35)$&\ \ 	0.2669	&\ \ 	1	&\ \ 	-0.3835	&\ \ 	0.6307	&\ \ 	0.4586	&\ \ 	0.8814	\\
$\Omega_mh^2$&\ \ 	0.3529	&\ \ 	-0.3835	&\ \ 	1	&\ \ 	-0.2549	&\ \ 	-0.0057	&\ \ 	0.07	\\
$\beta$&\ \ 	0.5802	&\ \ 	0.6307	&\ \ 	-0.2549	&\ \ 	1	&\ \ 	0.7176	&\ \ 	0.575	\\
$H(0.35) \,r_s(z_d)/c$&\ \ 	0.9259	&\ \ 	0.4586	&\ \ 	-0.0057	&\ \ 	0.7176	&\ \ 	1	&\ \ 	0.4981	\\
$D_A(0.35)/r_s(z_d)$&\ \ 	0.4832	&\ \ 	0.8814	&\ \ 	0.07	&\ \ 	0.575	&\ \ 	0.4981	&\ \ 	1	\\
\hline
\end{tabular}
\end{center}
\caption{Normalized covariance matrix of the measured and derived parameters, $\{H(0.35)$, $D_A(0.35)$, $\Omega_m h^2$, $\beta,$
$H(0.35) \,r_s(z_d)/c$, $D_A(0.35)/r_s(z_d)\}$ from SDSS DR7 LRGs using $\hat{\xi}_0+\hat{\xi}_2$. }
 \label{table:covar_matrix}
\end{table*}

\subsection{Comparison with Previous Work} 
\label{sec:compare}
While we have developed a general method to measure the dark energy and cosmological parameters that could be extracted from the galaxy 
clustering data alone, we restrict our method now by fixing some parameters to obtain the results
for comparison with previous work.

In our previous paper \citep{Chuang:2011fy}, we used full 2D correlation function and measured 
$H(z=0.35)=82.1_{-4.9}^{+4.8}$ km $s^{-1}$Mpc$^{-1}$ and $D_A(z=0.35)=1048_{-58}^{+60}$ Mpc, which are consistent with this study;
note that the full 2D correlation function captures more information than the leading multipoles.
\cite{Xu:2012fw} applied the density field reconstruction method on the same data 
and obtained $H(z=0.35)=84.4\pm7.1$ and $D_A(z=0.35)=1050\pm38$ Mpc, which are also in excellent agreement with our measurements.

\cite{Cabre:2008sz} measure $\beta$ from SDSS DR6 LRG using the normalized quadrupole defined by
Eq.(\ref{eq:Q}). To compare with their results, we make similar assumptions, and
use monopole-quadrupole method with fixing $\Omega_m=0.25$, $\Omega_b=0.045$, $h=0.72$, $n_s=0.98$, $k_{\star}=0.11$, 
and $\sigma_v=300$km/s in the $\Lambda$CDM model ($H(0.35)$ and $D_A(0.35)$ would also be fixed accordingly). 
Considering the scale range, $s=40-100h^{-1}$Mpc, we obtain $\beta=0.333\pm0.055$, in excellent agreement with
their measurement of $\beta=0.34\pm0.05$. Since the definition of the normalized quadrupole includes a integral of monopole with 
the minimum boundary from $s=0$, the advantage of using our effective multipole method instead of normalized quadrupole method is 
to avoid the distortion from the small scales where the scale dependent uncertainties are not well known. However, the 
distortion might be negligible compared to the statistical uncertainty of current measurements. 

\cite{Song:2010kq} split the same galaxy sample (SDSS DR7 LRG) to two redshift slices and obtained 
$\beta(z=0.25)=0.30_{-0.048}^{+0.047}$ and $\beta(z=0.38)=0.39\pm0.056$ without considering the 
geometric distortions. 
Their results are in excellent agreement with the values measured by \cite{Cabre:2008sz} and us
under the same assumptions. In addition, \cite{Blake:2011ep} measured $H(z)\,D_A(z)(1+z)/c$ = $0.28\pm0.04$ and $0.44\pm0.07$ at z = 0.22 and 0.41 respectively from WiggleZ survey \citep{Blake09, Parkinson:2012vd}. 
Linearly interpolating their results, we find the mean of $H(z)\,D_A(z)(1+z)/c$ to be 0.39 at z=0.35,
which is in excellent agreement with our measurement of $H(0.35)D_A(0.35)(1.35)/c$ = $0.38\pm0.06$.

\section{SYSTEMATICS} 
\label{sec:test}

Table.\ \ref{table:test} shows the systematic tests that we have
carried out by varying key assumptions made in our analysis.
These include the multipoles used, the range of scales used, 
the bin size used, and the minimum of the transverse separation used
to calculate the correlation function.

We use the results using $\hat{\xi}_0+\hat{\xi}_2$ as our fiducial results. We find that the constraints 
are stronger for using $\hat{\xi}_0+\hat{\xi}_2+\hat{\xi}_4$, but using $\hat{\xi}_0+\hat{\xi}_2+\hat{\xi}_6$ 
does not improve the constraints significantly. 
Therefore, it seems that $\hat{\xi}_0+\hat{\xi}_2+\hat{\xi}_4$ contains most of the information from the 2D 2PCF.
Since the measurements of $\hat{\xi}_0+\hat{\xi}_2+\hat{\xi}_4$ deviate from those of $\hat{\xi}_0+\hat{\xi}_2$ 
by about 1$\sigma$, we use the latter as our fiducial results to be conservative.

We vary the scale range chosen and the bin size used and find that the results are basically consistent. 
However, we find that the measurement of $D_A(z)/r_s(z_d)$ is more stable than that of $H(0.35) \,r_s(z_d)/c$. It might indicate the appearance 
of systematic errors from the measurement of the correlation function in the direction along the line of sight. 
While the observed correlation function along the line of sight is noisier and harder to model due to galaxy perculiar velocities,
we test the impact of systematic uncertainties along the line of sight by removing the data with the transverse separation, $\sigma$, 
smaller than 5 or 10 $h^{-1}$Mpc. We find that the results are insensitive to this.
Thus our measurement of $H(0.35) \,r_s(z_d)/c$ should not be contaminated by systematic errors along the line of sight.

There is possible systematic uncertainty from the radial selection function used to construct the random catalogs.
\cite{Ross:2012qm} found that the least biased way is using
"shuffled" method for SDSS-III/BOSS DR9 CMASS sample. Shuffled method is to assign the redshift of a galaxy of the random catalog with the redshift of the observed data picked randomly.
\cite{Samushia:2012iq} found that using spline method, which is the same method as we use in this study,
could obtain less biased result for SDSS DR7 LRG sample. In fact, the biased effect due to the
radial selection function depends on the galaxy sample, survey geometry,
and scale range studied. For example, for a narrow beam survey,
while most of the structure is in the line of sight direction,
shuffled method would erase most of the information. \cite{Samushia:2012iq}  
showed that the spline method has least bias for the SDSS DR7 LRG
sample in the scale range we are interested ($s > 40h^{-1}$Mpc). And the
bias is much smaller then the statistic error. Therefore, we expect the bias to be negligible.

\begin{table*}\scriptsize
\begin{center}
\begin{tabular}{crrrrrr}
\hline

&$H(0.35)$ &$D_A(0.35)$  &$\Omega_m h^2$ &$\beta$ &$H(0.35) \,r_s(z_d)/c$ &$D_A(0.35)/r_s(z_d)$ \\ \hline

$\hat{\xi}_0+\hat{\xi}_2$ (fiducial)	
&\ \ $	80.0	_{	-8.6	}^{+	8.2	}$
&\ \ $	1063	_{	-85	}^{+	87	}$
&\ \ $	0.103	\pm0.015$
&\ \ $	0.45	_{	-0.14	}^{+	0.15	}$
&\ \ $	0.0437	_{	-0.0043	}^{+	0.0041	}$
&\ \ $	6.48	_{	-0.43	}^{+	0.44	}$\\

$\hat{\xi}_0+\hat{\xi}_2+\hat{\xi}_4$
&\ \ $	87.3	\pm6.4$
&\ \ $	1095	\pm58$
&\ \ $	0.107	_{	-0.014	}^{+	0.015	}$
&\ \ $	0.54	\pm0.11$
&\ \ $	0.0472	\pm0.0031$
&\ \ $	6.75	_{	-0.23	}^{+	0.24	}$\\

$\hat{\xi}_0+\hat{\xi}_2+\hat{\xi}_6$
&\ \ $	78.5	_{	-8.9	}^{+	8.7	}$
&\ \ $	1025	_{	-82	}^{+	88	}$
&\ \ $	0.107	\pm0.016$
&\ \ $	0.41	\pm0.14$
&\ \ $	0.0424	_{	-0.0043	}^{+	0.0044	}$
&\ \ $	6.31	\pm0.49$\\

$30<s<120$
&\ \ $	85.1	_{	-8.2	}^{+	7.8	}$
&\ \ $	1072	_{	-62	}^{+	64	}$
&\ \ $	0.115	\pm0.014$
&\ \ $	0.38	\pm0.10$
&\ \ $	0.0453	_{	-0.0039	}^{+	0.0037	}$
&\ \ $	6.71	_{	-0.30	}^{+	0.31	}$\\

$50<s<120$
&\ \ $	77.5	_{	-8.4	}^{+	8.2	}$
&\ \ $	1034	_{	-109	}^{+	103	}$
&\ \ $	0.101	_{	-0.018	}^{+	0.019	}$
&\ \ $	0.5	_{	-0.18	}^{+	0.19	}$
&\ \ $	0.0425	_{	-0.0040	}^{+	0.0038	}$
&\ \ $	6.27	_{	-0.61	}^{+	0.55	}$\\

$40<s<110$
&\ \ $	73.5	_{	-7.0	}^{+	6.7	}$
&\ \ $	1064	_{	-76	}^{+	77	}$
&\ \ $	0.107	\pm0.015$
&\ \ $	0.35	\pm0.11$
&\ \ $	0.0398	_{	-0.0034	}^{+	0.0031	}$
&\ \ $	6.54	_{	-0.40	}^{+	0.41	}$\\

$40<s<130$
&\ \ $	83.5	_{	-8.7	}^{+	8.3	}$
&\ \ $	1082	_{	-75	}^{+	78	}$
&\ \ $	0.105	\pm0.015$
&\ \ $	0.48	\pm0.15$
&\ \ $	0.0454	_{	-0.0043	}^{+	0.0041	}$
&\ \ $	6.63	_{	-0.35	}^{+	0.37	}$\\

bin size = 4 $h^{-1}$Mpc
&\ \ $	77.3	_{	-7.2	}^{+	7.1	}$
&\ \ $	1026	_{	-85	}^{+	86	}$
&\ \ $	0.113	\pm0.017$
&\ \ $	0.39	\pm0.12$
&\ \ $	0.0413	\pm0.0034$
&\ \ $	6.39	\pm0.45$\\

bin size = 8 $h^{-1}$Mpc
&\ \ $	79.1	_{	-8.6	}^{+	8.1	}$
&\ \ $	1065	_{	-78	}^{+	84	}$
&\ \ $	0.112	\pm0.017$
&\ \ $	0.42	\pm0.13$
&\ \ $	0.0423	_{	-0.0041	}^{+	0.0039	}$
&\ \ $	6.62	_{	-0.37	}^{+	0.41	}$\\

$\sigma > 5 h^{-1}$Mpc
&\ \ $	79.6	_{	-8.4	}^{+	8.3	}$
&\ \ $	1054	_{	-85	}^{+	87	}$
&\ \ $	0.103	\pm0.015$
&\ \ $	0.43	\pm0.14$
&\ \ $	0.0434\pm0.0042$
&\ \ $	6.43	_{	-0.42	}^{+	0.45	}$\\

$\sigma > 10 h^{-1}$Mpc
&\ \ $	76.5	_{	-7.8	}^{+	7.5	}$
&\ \ $	1048	\pm	84	$
&\ \ $	0.099	\pm	0.014$
&\ \ $	0.37	\pm	0.13	$
&\ \ $	0.0420	_{	-0.0039	}^{+	0.0036	}$
&\ \ $	6.34	\pm	0.44	$\\

\hline
$\frac{mean_{max}-mean_{min}}{\left(\sigma^+_{fid}+\sigma^-_{fid}\right)/2}$
&\ \  1.64
&\ \ 0.81
&\ \ 1.07
&\ \ 1.31
&\ \ 1.76
&\ \ 1.01\\
\hline
\end{tabular}
\end{center}
\caption{This table shows the systematic tests that vary the
combination of multipoles, the scale range, the bin size, and the minimum transverse separation used in the analysis.
The fiducial results are obtained using $\hat{\xi}_0+\hat{\xi}_2$, the scale range $40<s<120\ h^{-1}$Mpc, a bin size of $5h^{-1}$Mpc, and no minimum transverse separation.
The other results are calculated with only specified quantities different from the fiducial one.
The unit of $H$ is $\Hunit$. The unit of $D_A$ and $r_s(z_d)$ is $\rm Mpc$. In the last row, we show the variation between these tests 
by computing the maximum difference between the mean values divided by the errors of the fiducial measurements.
} \label{table:test}
\end{table*}

\section{Conclusion and Discussion} 
\label{sec:conclusion}

We have demonstrated the feasibility of using multipoles of the correlation function to measure $H(z)$, $D_A(z)$, $\Omega_mh^2$, and $\beta$
by applying the method to individual mock catalogs from LasDamas in an MCMC likelihood analysis.

The method we developed is modified from \cite{Chuang:2011fy}, which was the first method to include the geomatric distortion 
(also known as Alcock-Paczynski test, see \cite{Alcock:1979mp}) on galaxy clustering data on large scales. We compute the multipoles 
from the theoretical and observed 2D 2PCF in the same way, thus the only approximation made is that
the distance of any pair of galaxies can be converted with two stretch factors between 
different models in the redshift range considered.

We have obtained the constraints for 
the measured and derived parameters, $\{H(0.35)$, $D_A(0.35)$, $\Omega_m h^2$, $\beta$,
$H(0.35) \,r_s(z_d)/c$, $D_A(0.35)/r_s(z_d)\}$,
from the multipoles of the correlation function from the sample of SDSS DR7 LRGs  
which are summarized by Tables\ \ref{table:mean_sdss_quad} and \ref{table:mean_sdss_four}.

We find that while the mean values of estimated parameters remain stable 
(with rare deviations) for the mock data when higher multipoles are used, this is not true for the
SDSS DR7 LRG data. We find $H(0.35)\,r_s(z_d)/c$ = $0.0437_{-0.0043}^{+0.0041}$ 
using monopole + quadrupole, and $H(0.35)\,r_s(z_d)/c$ = $0.0472\pm0.0031$ 
using monopole + quadrupole + hexadecapole. 
This deviation could be caused by
statistical variance. 
In addition, there is some deviation between the LasDamas measurments and the theoretical model for hexadecapole.
However, the deviation is small compared to the uncertainties of the measurements.
To be conservative, we choose the measurement using 
monopole + quadrupole as our fiducial results.

\section*{Acknowledgements}
We are grateful to the LasDamas project              
for making their mock catalogs publicly available.
The computing for this project was performed at the OU 
Supercomputing Center for Education and Research (OSCER) at the University of 
Oklahoma (OU).
This work used the Extreme Science and Engineering Discovery Environment (XSEDE), which is supported by National Science Foundation grant number OCI-1053575. 
This work was supported by DOE grant DE-FG02-04ER41305.
C.C. was also supported by the Spanish MICINN’s Consolider-Ingenio 2010 Programme under grant MultiDark CSD2009-00064 and grant AYA2010-21231,
and by the Comunidad de Madrid under grant HEPHACOS S2009/ESP-1473.

Funding for the Sloan Digital Sky Survey (SDSS) has been provided by the Alfred P. Sloan Foundation, the Participating Institutions, the National Aeronautics and Space Administration, the National Science Foundation, the U.S. Department of Energy, the Japanese Monbukagakusho, and the Max Planck Society. The SDSS Web site is http://www.sdss.org/.

The SDSS is managed by the Astrophysical Research Consortium (ARC) for the Participating Institutions. The Participating Institutions are The University of Chicago, Fermilab, the Institute for Advanced Study, the Japan Participation Group, The Johns Hopkins University, Los Alamos National Laboratory, the Max-Planck-Institute for Astronomy (MPIA), the Max-Planck-Institute for Astrophysics (MPA), New Mexico State University, University of Pittsburgh, Princeton University, the United States Naval Observatory, and the University of Washington. 
\setlength{\bibhang}{2.0em}

\label{lastpage}

\end{document}